\documentclass{article}

\usepackage{arxiv}

\usepackage[utf8]{inputenc} 
\usepackage[T1]{fontenc}    
\usepackage{hyperref}       
\usepackage{url}            
\usepackage{booktabs}       
\usepackage{amsfonts}       
\usepackage{nicefrac}       
\usepackage{microtype}      
\usepackage{lipsum}
\usepackage{graphicx}
\graphicspath{ {./images/} }
\usepackage{xcolor}
\usepackage{dirtytalk}
\usepackage{subcaption}
\usepackage{multirow}
\usepackage{amsmath}
\usepackage[export]{adjustbox}

\usepackage{authblk}

\usepackage{color, colortbl}
\definecolor{LightCyan}{rgb}{0.88,1,1}
\definecolor{Gray}{gray}{0.8}
\definecolor{LightGray}{gray}{0.95}

\usepackage[english]{babel}
\usepackage{amsthm}

\theoremstyle{definition}
\newtheorem{definition}{Definition}[section]

\usepackage{caption}

\makeatletter
\def\blfootnote{\xdef\@thefnmark{}\@footnotetext}
\makeatother

\newcounter{noteDMctr} 

\newcounter{noteMCctr} 

\definecolor{colour3}{RGB}{178,55,250} 
\newcommand{\mc}[1]{\textcolor{black}{{{}}#1}}

\newcommand{\mcc}[1]{\textcolor{black}{{{}}#1}}

\newcommand{\dm}[1]{\textcolor{black}{{{}}#1}}
\newcommand{\dhm}[1]{\textcolor{black}{{{}}#1}}

\usepackage[nodayofweek,level]{datetime}
\newcommand{\mydate}{\formatdate{21}{12}{2022}}

\title{DeFi: data-driven characterisation of Uniswap v3 ecosystem \& an ideal crypto law for liquidity pools}

\author[1, 4, *]{Deborah Miori}
\author[1, 2, 3, 4]{Mihai Cucuringu}
\affil[1]{Mathematical Institute, University of Oxford, Oxford, UK}
\affil[2]{Department of Statistics, University of Oxford, Oxford, UK}
\affil[3]{The Alan Turing Institute, London, UK}
\affil[4]{Oxford-Man Institute of Quantitative Finance, Oxford, UK}
\affil[*]{\text{Corresponding author: Deborah Miori,} \url{deborah.miori@maths.ox.ac.uk}}
\begin{document}
\maketitle
\begin{abstract}
\dhm{Uniswap is a Constant Product Market Maker built around liquidity pools, where pairs of tokens are exchanged subject to a fee that is proportional to the size of transactions. At the time of writing, there exist more than 6,000 pools associated with Uniswap v3, implying that empirical investigations on the full ecosystem can easily become computationally expensive. Thus, we propose a systematic workflow to extract and analyse a meaningful but computationally tractable sub-universe of liquidity pools.} 
\dhm{Leveraging on the $34$ pools found relevant for the six-months time window January-June 2022, we then investigate the related liquidity consumption behaviour of market participants.} 
\dhm{We propose to represent each liquidity taker by a suitably constructed}
\textit{transaction graph}, \dhm{which is a fully connected network where nodes are the liquidity taker's executed transactions,}
and edges \mcc{contain weights encoding} the time elapsed between any two transactions.  
We extend the NLP-inspired \textit{graph2vec} algorithm to
the weighted undirected setting, 
and employ it to obtain an embedding of the set of graphs. 
\dhm{This embedding allows us to extract seven clusters of liquidity takers, with equivalent behavioural patters and interpretable trading preferences.} 
\dhm{We conclude our work by testing for relationships between the characteristic mechanisms of each pool, i.e. liquidity provision, consumption, and price variation. We introduce a related } \textit{ideal crypto law}, \dhm{inspired from the ideal gas law of thermodynamics, and demonstrate that pools adhering to this law are healthier trading venues in terms of sensitivity of liquidity and agents' activity. Regulators and practitioners could benefit from our model by developing related pool health monitoring tools.}  
\end{abstract}


\section{Introduction}
\label{sec:intro}

A \textit{blockchain} is a type of Distributed Ledger Technology (DLT) that stores users transactions on an increasingly long sequence of blocks of data. The ledger is \dhm{replicated across a network of servers}
to allow the validation of new transactions by the peer-to-peer (P2P) computer network and 
\dhm{consequent}
addition of blocks, \dhm{thus increasing trust, security, transparency, and the traceability of data.}
\dhm{Bitcoin was the first blockchain to acquire worldwide notoriety. It was designed by the person(s) known via the pseudonymous Satoshi Nakamoto during 2007 and 2008, and subsequently described in the whitepaper \cite{nakamoto2009bitcoin} in 2009.}
The project was \dhm{indeed} released as an open source software in 2009 and from that moment, 
Bitcoin started slowly acquiring increasing value and seeing higher trading volumes. However, Bitcoin 
lacks the capacity for logical loops and conditionals, \dhm{and this} limitation fueled the rise of the Ethereum blockchain,  which was first described in the 2013 whitepaper   \cite{Buterin2013}.

Ethereum supports smart contract functionality and, due to this, it is able to offer financial \dhm{services}
that do not rely on intermediaries such as brokerages, exchanges or banks. Thus, Ethereum \dhm{is commonly considered as the protocol that first allowed the formulation of foundations} 
for Decentralised Finance (DeFi). 
Within DeFi, individuals can lend, trade, and borrow  using software that automatically broadcasts their  intentions for P2P verification. \dhm{Then, valid financial actions are recorded on the blockchain.} 
Decentralised  Exchanges (DEXs) are a direct result of this setup, and started being designed and implemented mainly from 2017. They differ from the usual centralised exchanges, since they are non-custodial and leverage the self-execution of smart contracts for P2P trading, allowing users to retain control of their private keys and funds. One of the first and most established DEXs at the time of writing is Uniswap, built on Ethereum and launched in November 2018. There exist three versions of Uniswap (namely v1, v2, v3, see the whitepapers \cite{whitepaperV1}, \cite{uniswap-v2}, \cite{uniswap-v3} respectively) that update its design and evolve its functionalities. 

We focus on Uniswap v3 data in this research, and 
investigate the related ecosystem. \dhm{Thus, we now highlight some of Uniswap core aspects and begin with the simpler features that characterise its first two versions v1 and v2.}
Uniswap is an Automated Market Maker (AMM), and in particular, a Constant Function Market Maker (CFMM).  This implies that digital assets are traded without centralised permission and the pricing occurs following a mathematical formula, rather than relying on an order book as in traditional exchanges. Uniswap smart contracts hold liquidity reserves of various token pairs and trades initiated by liquidity takers (LTs) are executed directly against these reserves. Reserves are pooled \dhm{from} a network of liquidity providers (LPs) who supply the system with tokens in exchange for a proportional share of transaction fees.  
A standard Uniswap liquidity pool allows the exchange, or \textit{swap}, of two assets via the constant product market maker mechanism
\begin{equation}
(x-\Delta x) \cdot \Big[ y + \big(1-\frac{\gamma}{10^6}\big) \Delta y \Big] = x \cdot y = k, 
\label{eq:AMM}
\end{equation}  
where $x,y \in \mathbb{Q}^+$ are the current pool reserves of tokens $X,Y$ respectively. Then, $k \in \mathbb{Q}^+$ tracks the evolution of liquidity of the pool, and $\gamma \in \{100, 500, 3000, 10000 \}$ (i.e. $\{1, 5, 30, 100 \}$ basis points) denotes the \textit{feeTier} characteristic of the pool.
Here, a LT sells an amount $\Delta y$ of token $Y$ to the pool and receives $\Delta x$ of token $X$ back. The instantaneous exchange rate $Z$ between the two digital assets is given by the proportion of respective reserves in the pool, i.e. 
\begin{equation}
    Z=\frac{x}{y},
\label{exchange-rate}
\end{equation}
and changes following the trades of LTs. A proportion $\frac{\gamma}{10^6}$ of each swap is kept by the protocol to reward LPs for their service. Swaps do not change $k$, while this invariant does vary if new liquidity is \textit{minted} (added) or \textit{burned} (withdrawn) in the pool by LPs. Of course, higher liquidity assures less price slippage for LTs and is thus preferred. LPs profit an amount proportional to their involvement into the whole liquidity of the pool for each trade occurred. However, they also incur \textit{impermanent loss} due to the need to stake both tokens to provide liquidity, while bearing the risk of varying exchange rates. 

In Uniswap v3, \dhm{the main difference from the previous versions is that} \textit{concentrated liquidity} is implemented. \dhm{The continuous space of exchange rates of pools is discretised into intervals whose boundaries are called \textit{ticks}, and every minting action of LPs specifies two ticks between which the liquidity will specifically be provided.}
This means that LPs can choose the range of prices and proportions over which they \dhm{lock} their tokens. However, concentrated liquidity also implies that LPs collect LTs' fees only while the exchange rate of executed trades lies between two ticks over which they are indeed providing liquidity. \dhm{For a full characterisation of the wealth of LTs and LPs in Uniswap v3, we point to the mathematical analyses pursued in \cite{cartea2,cartea1}.} 

For completeness, it is also worth mentioning that every action (i.e. creation of a pool, swap, mint or burn operation...) that occurs on Uniswap, or in the general DeFi universe, must be validated and registered on the blockchain to be considered executed. This introduces a further cost for the initiator of the action, who needs to pay non-negligible \textit{gas fees} \cite{eth-txns-fees} to miners to compensate them for the computational power they \dhm{devote}. This is especially significant for blockchains that use a Proof-of-Work  consensus protocol, such as Ethereum until September 2022. Indeed, Ethereum transitioned to Proof-of-Stake via the upgrade named  \say{The Merge}, allowing validation of transactions to not only rely on computational power, and opening the opportunity to have lower gas fees and enhanced users participation in DeFi (despite not yet reality). 

Many further interesting new and old finance concepts live within DeFi and beyond DEXs. One such concept worth mentioning first is that of  \textit{stablecoins}. Stablecoins are digital assets that are pegged to the value of a fiat currency, and can be useful to exit risky positions while remaining inside the crypto ecosystem. 
Some stablecoins are fiat-backed (e.g. USDC, Tether), while others are backed by an over-collateralised pool of cryptocurrencies (e.g. DAI). There also exist algorithmic coins (e.g. UST), which closely resemble traditional pegged exchange rates and are indeed also vulnerable to speculative attacks, i.e. as it happened with the Terra-Luna crash in May 2022. Apart from stablecoins, DeFi provides several lending protocols (e.g. Aave, Compound, Instadapp, Maker), protocols for derivatives trading (e.g. dYdX, Futureswap, Nexus), and DEX aggregators (e.g. 1inch) that optimise routing to take advantage of the best exchange rates across multiple other exchanges. In \cite{disentangling}, there is an interesting study of the interactions between different blockchain protocols.

While DeFi is fascinating, it is also the stage of many scams, speculative high-risk  investments, direct blockchain attacks, and money laundering events. On top of that, its complexity and atomicity might disadvantage small users, whose transactions can e.g. be re-ordered before execution by the validators for their own profit, known here as \textit{miner extractable value}  (MEV). Despite the current effort of regulators to penetrate the crypto world and establish some equilibrium between centralisation and decentralisation \dhm{of power}, the current situation and possible \mc{upcoming developments are still highly confusing, especially for outsiders or newcomers}. Interesting overviews and critical thoughts are presented in \cite{defi-schar} and \cite{defi-overview}, where the latter work especially discusses enforcing tax compliance, anti-money laundering laws and how to possibly prevent financial malfeasance. 
The different layers of DeFi are studied in \cite{managing-risk-defi-portfolios}, where the related specific risks, i.e. at the blockchain, protocol, pool, and token level, are also analysed \dhm{and a risk parity approach for portfolio construction in Uniswap is proposed.} 

The current academic research is also at its early stages in terms of understanding the inner dynamics of DEXs and external relationships with the well-known traditional stock market, especially from an empirical and data-driven point of view. 
\dhm{Among interesting recent studies there is} \cite{LP-contract-design}, where the authors investigate how promoting a greater diversity of price-space partitions in Uniswap v3 can simultaneously benefit both liquidity providers and takers. Then, \cite{econAMM} studies whether AMM protocols, such as Uniswap, can sustainably retain a portion of their trading fees for the protocol and the expected outflow of traders to competitor venues.  Inefficiencies between Uniswap and SushiSwap are investigated in \cite{ineff-Uniswap}, where sub-optimal trade routing is addressed. However, \cite{analysis-uniswap} shows that constant product markets should closely track the reference market price.
Flows of liquidity between Uniswap v2 pools are studied in \cite{behaviour-LP}, while \cite{heimbach2022risks} shows the difficulty of earning significant returns by providing liquidity in Uniswap v3. \dhm{Interestingly, \cite{cartea2} fully characterises the wealth of LTs and LPs in Uniswap v3, and shows that LPs suffer a \say{Predictable Loss}.}


\paragraph{Main Contributions.} We divert from the available literature in many ways. 
To the best of our knowledge, this is the first study to systematically  \dhm{identify}
a set of pools that are necessary to be considered for a full view of Uniswap v3 dynamics. We assess both the pools' inner features and their interconnectedness, and 
\dhm{present} a workflow for extracting significant sub-universes of pools in time, which can be completely reproduced by the reader. 
Then, we leverage on this first point to cluster LTs and characterise their broad behaviour. 
Our final contribution is to propose an \textit{ideal crypto law} for liquidity pools, inspired by the ideal gas law from thermodynamics. We provide motivation for it and show that pools with high \textit{cryptoness}, i.e. strongly adhering to \dhm{the proposed} law, are healthier crypto environments on which to trade, \dhm{by their levels of liquidity and activity of agents}. The \dhm{strength} of  \textit{cryptoness} of a pool  
can evolve in time and hence it is important to track it, along with various metrics that quantify the risks associated to the respective pool.

\paragraph{Structure of the Paper.} In Section  \ref{sec:pools-of-interest}, we identify the most important and interconnected liquidity pools for different time windows within 2022. Next, we cluster LTs according to their behaviour on the relevant sub-universes in Section \ref{sec:cluster}. Due to the complexity of this ecosystem, we draw on intuition from Natural Language Processing (NLP) and graph embedding techniques to assess structural equivalence of trading behaviour in a novel way. Section \ref{sec:physics} expands our investigations by proposing an \textit{ideal crypto law} to simultaneously model LTs, LPs and price dynamics for each pool under consideration. Finally, we summarise our thoughts and discuss future research directions in Section \ref{sec:conclusions}.


\section{Systematic selection of Uniswap v3 pools of interest}
\label{sec:pools-of-interest}

\subsection{Empirical introduction to the ecosystem}

\begin{figure}[t]
    \centering
    \includegraphics[width=0.65\linewidth]{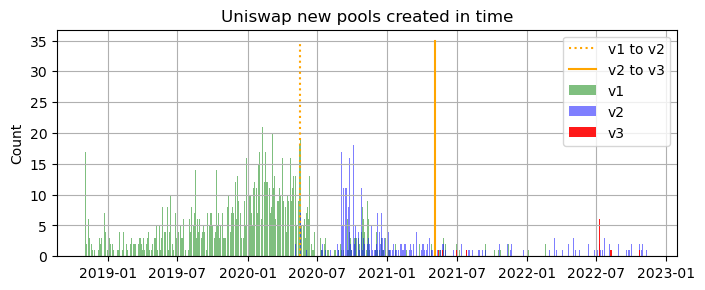}
    \caption{Daily count of new pools created via UniswapV1Factory, UniswapV2Factory and UniswapV3Factory smart contracts. The two vertical orange lines depict the dates of official transition from Uniswap v1 to v2, and from v2 to v3.}
    \label{fig:new-pools-time}
\end{figure}

At the time of writing, Uniswap v3 is the latest implementation of this DEX. It launched in May 2021, introduced the concept of concentrated liquidity and allowed multiple feeTiers. For each Uniswap version $N=1,2,3$, the addresses of related liquidity pool smart contracts are stored in the respective \say{UniswapV$N$Factory} contracts. We access them on Etherscan\footnote{\url{https://etherscan.io/}} by querying for transactions related to UniswapV$N$Factory addresses\footnote{UniswapV1Factory: 0xc0a47dFe034B400B47bDaD5FecDa2621de6c4d95}\footnote{UniswapV2Factory: 0x5C69bEe701ef814a2B6a3EDD4B1652CB9cc5aA6f}\footnote{UniswapV3Factory: 0x1f98431c8ad98523631ae4a59f267346ea31f984} and filtering for methods \say{Create Exchange}, \say{Create Pair} or \say{Create Pool}, for the three versions respectively. We find total numbers of $3,857$ and $992$ and $40$ respectively associated calls until 15 November 2022, which we agglomerate at daily level and plot in Fig. \ref{fig:new-pools-time}. The dates of transition from Uniswap v1 to v2, and v2 to v3, are also depicted. It is interesting to notice that the previous protocols remain active after the transitions, but their liquidity can be easily moved to the new Uniswap versions via \say{Migrator} contracts. In terms of pools, the main differences between v1, v2 and v3 are that the first protocol allows only pools where one token is ETH and feeTier $\gamma=3000$, while the second one introduces the ability to create a pool between any two tokens. Then, v3 expands pools to feeTiers $100$, $500$ or $10000$.
Although there is a total of $4,889$ pools directly created with UniswapV$N$Factory contracts, the majority of them is the result of the 2020-21 cryptomania and inflated creation of new tokens.
This translates to many pools not containing any relevant amount of liquidity locked, but which do not disappear due to the immutability of the blockchain. On the other hand, we also expect wrapped calls to the Factory contracts and thus refer to the above as a lower bound to the number of pools created.

\begin{figure}[h]
    \centering
    \includegraphics[width=0.65\linewidth]{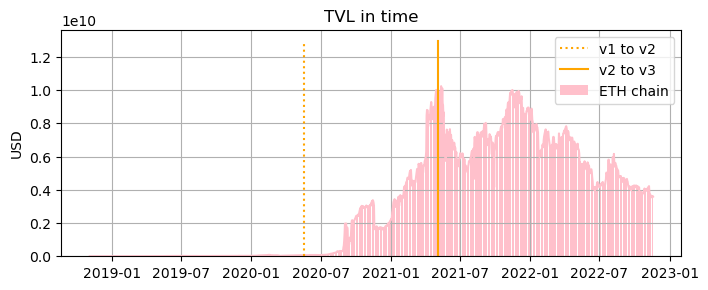}
    \caption{Evolution in time of the TVL in USD on Uniswap main Ethereum chain. The higher the TVL, the more liquid the ecosystem is considered to be. In orange, we show the dates of transition from Uniswap v1 to v2, and from v2 to v3.}
    \label{fig:TVL-time}
\end{figure}

As a last note, we show the evolution of Uniswap liquidity on its main Ethereum chain in Fig. \ref{fig:TVL-time}. The data are downloaded via the Defi Llama\footnote{https://defillama.com/} API and we proxy liquidity by the total amount of USD locked on the protocol, i.e. Total Value Locked (TVL).

\subsection{Data download and coarse refinement of pools}

Each version of Uniswap has its own dedicated subgraph, which has a precise endpoint for querying data and a database schema to expose the available fields. Our terminology follows the Uniswap v3 schema\footnote{\url{https://github.com/Uniswap/v3-subgraph/blob/main/schema.graphql\#L1}} and we download pools data via the related subgraph.
Our first aim is to identify the pools most representative of the Uniswap ecosystem, which we interpret as having significant liquidity consumption and provision events, but also showing high interconnectedness. To this end, we download the latest summary data of all possible pools, full historical record of liquidity consumption operations, and full record of liquidity provision actions. Then, we develop a systematic approach that aims at increasingly discarding layers of pools with weakest features first and then weakest dynamics too, respectively in this subsection and in the next one. The final data set will be our starting point for the subsequent analyses, but aims at being useful to a wider group of researchers that desire to empirically investigate Uniswap v3.

\paragraph{Download summary data of pools.}
As of 15 November 2022, we download the latest \say{Pool} data as described in Uniswap v3 subgraph schema.
We download pools in descending transaction count (txnCount) order, since this variable is strictly increasing in time, while e.g. TVL is not. This allows us to select a universe of pools which have had a minimum number of transactions thus far. 
We apply this weak initial filtering on the first $6,000$ pools by txnCount, and find that only $1,344$  pools report at least $1,000$ transactions by 15 November 2022.  
Then, we also restrict ourselves to pools where both exchanged tokens are traded in at least $3$ pools (e.g. token $T$ is  traded against a stablecoin, against ETH and against ETH with different feeTier), in order to focus on  interesting dynamics of the full ecosystem.  The result is that we subset to a universe of $696$ pools to consider in our study. 

\paragraph{Download LP data.} We download  liquidity provision data for these $696$ pools and find non-empty entries for $629$ of them. Liquidity provision data are downloaded to have a historical record of all liquidity mint and burn operations on each pool, with related USD value.
By computing the total cumulative sum of LPs activity, we proxy the TVL in USD that each pool contains at every moment in time and denote it as \say{proxyTVL}. Unfortunately, we cannot simply rely on the \say{PoolDayData} values provided by the subgraph due to incoherences found when cross-checking with Ethereum blochckain data on Etherscan. 

Uniswap v3 data start on 6 May 2021, when the transition from the previous version of the protocol successfully completed. While for the first months only the $500$, $3000$ and $10000$ feeTiers were implemented, in November 2021 a fourth feeTier $\gamma = 100$ was activated. This generated structural flows, noise and adjustments that we prefer to exclude from our analyses. On top of that, we recognise that the transition of Uniswap's foundation blockchain (i.e. Ethereum) from Proof-of-Work to Proof-of-Stake in September 2022 could have triggered  turbulences on the ecosystem too.
Thus, we decide to focus our analyses on the six-months period from January 2022 to the end of June 2022, which we consider as  the most representative of the actual DEX dynamics. We check for every pool if our proxyTVL passes the threshold of $1,000,000$ USD (one million dollars) at any point before the end of June 2022. This is motivated by the aim to find pools that were liquid enough at some point in our time window to capture interesting behaviour of LTs and LPs.
We find $282$ pools that satisfy this further requirement. Of these, $210$ had that much TVL already at some point before January 2022, and $261$ at some point before April 2022.
While some pools acquire relevance as time passes, other ones can also lose liquidity, as in the extreme case of pools related to the Terra-Luna crash of May 2022.
As a final detail, we highlight that we additionally check whether a pool has at least one million USD in TVL for two consecutive points in time in order to avoid pools where a substantial amount of liquidity is minted and immediately burned by an agent, to likely take advantage of specific external information.

\begin{figure}[h]
    \centering
    \includegraphics[width=0.75\linewidth]{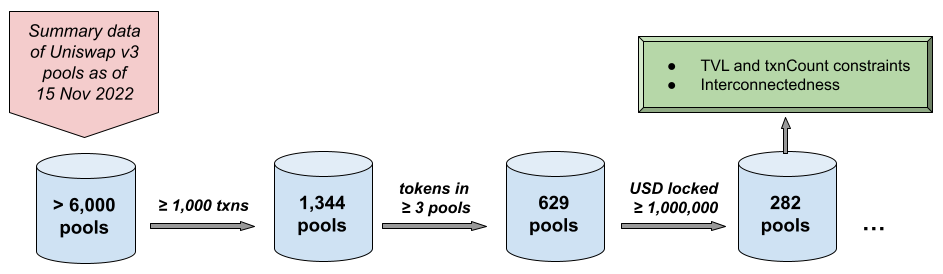}
    \caption{Summary diagram of the filtration steps pursued during our coarse refinement of pools. Stronger constraints on the TVL and txnCount of pools will follow in the next subsection, such as an attention to maximise the interconnectedness of the final sub-universe of pools.}
    \label{fig:diagram-filtration}
\end{figure}

\paragraph{Download LT data.} For the above $282$ pools, we download related liquidity consumption data and find all non-empty data sets. Thus, we are left with a final set of $282$ liquidity pools, for which we have a summary file, a LP database, and a LT database each. This completes our coarse filtering of pools, which implements the least invasive possible initial requirements, while still filtering down  the universe of Uniswap v3 liquidity pools to a tractable number of instances. The diagram in Fig. \ref{fig:diagram-filtration} summarises the steps completed.

\subsection{Final refinements}

\paragraph{Stronger TVL and txnCount constraints.} From our first coarse filtration, we recover $282$ pools to consider further in Uniswap v3 analyses. However, we shall be more strict about the minimum number of transactions taking place on a pool and its TVL in time, in order to lower the noise-to-signal ratio in the data. 
As already motivated, we wish to focus on the six-months window $[January, July)$ 2022, which we denote as our case $A$. We also consider five sub-ranges, namely the two three-months windows $[January, April)$, $[April, July)$ that we denote as cases $B1/B2$, and the three two-months windows $[January, March)$, $[March, May)$, $[May, July)$, that we call cases $C1/C2/C3$.
For each case and related time window $[start, end)$, we extract the pools with at least $1,000$ transactions before $start$ (where the number of transactions in time is calculated via the cumulative sum of both swap events and mint or burn operations) and that also had at least $1,000,000$ USD in proxyTVL both at the $start$ and $end$ of the interval.
Considering sub-ranges allows us to further account for the appearance of new pools that became significantly liquid or active after January 2022, or pools that lost the majority of their liquidity before July 
\dhm{2022, thus lowering survivorship/loser bias.}
For cases $A/B1/B2/C1/C2/C3$ in order, we find respectively $113/126/148/131/146/155$ pools that satisfy the above requirements, for which we save the related addresses and information.
Taking the union of these sets of pools, we notice that we are considering $177$ different pools overall. Of these, five pools belong to the $100$ feeTier, $28$ pools to the $500$ feeTier, $84$ pools to the $3000$ feeTier and $60$ pools to the $10000$ feeTier.

\begin{figure}[t]
    \begin{subfigure}{.5\textwidth}
      \centering
      \includegraphics[width=.99\linewidth]{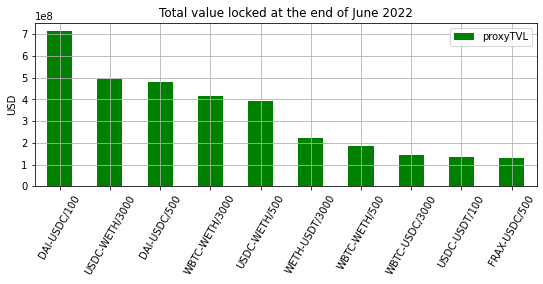}
      \caption{Pools with highest liquidity at the end of June 2022.}
      \label{Barplots-info-tvl}
    \end{subfigure}%
    \begin{subfigure}{.5\textwidth}
      \centering
      \includegraphics[width=.99\linewidth]{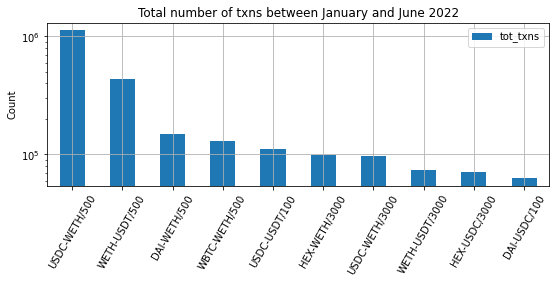}
      \caption{Pools with highest total number of transactions during case $A$.}
      \label{Barplots-info-txn}
    \end{subfigure}
    \caption{The $10$ most liquid and active pools for case $A$, i.e. over the time window between January and June 2022.}
    \label{Barplots-info}
\end{figure}

\begin{figure}[h]
    \centering
    \begin{subfigure}{.5\textwidth}
      \centering
      \includegraphics[width=.8\linewidth]{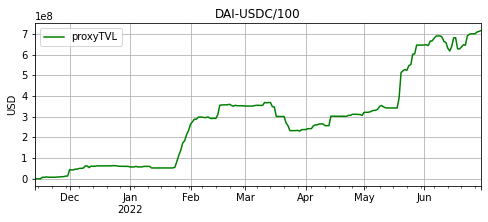}
      \caption{Full history of TVL for the pool with highest final TVL.}
      \label{fig:sub1}
    \end{subfigure}%
    \begin{subfigure}{.5\textwidth}
      \centering
      \includegraphics[width=.83\linewidth]{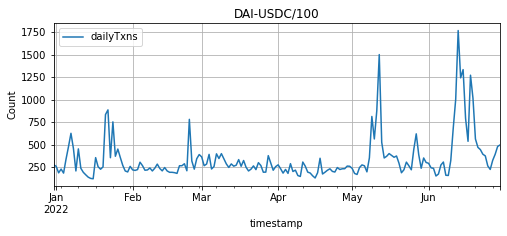}
      \caption{Daily txns count for the pool with highest final TVL.}
      \label{fig:sub2}
    \end{subfigure}
\vspace{0.5cm}
\begin{subfigure}{.5\textwidth}
      \centering
      \includegraphics[width=.8\linewidth]{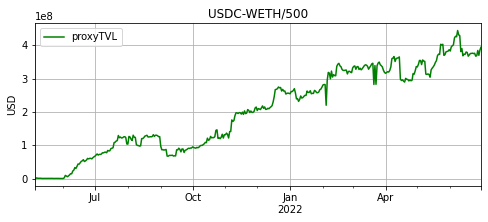}
      \caption{Full history of TVL for the pool with highest total txnCount.}
      \label{fig:sub1}
    \end{subfigure}%
    \begin{subfigure}{.5\textwidth}
      \centering
      \includegraphics[width=.83\linewidth]{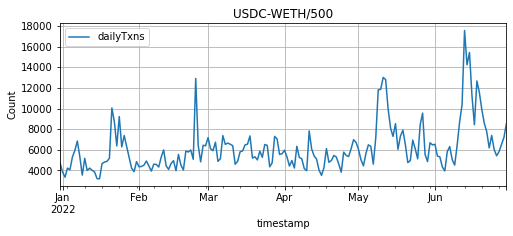}
      \caption{Daily txns count for the pool with highest total txnCount.}
      \label{fig:sub2}
    \end{subfigure}
\caption{Evolution of liquidity, i.e. proxyTVL, and daily number of transactions for the pool with highest TVL at the end of June 2022 (DAI-USDC/100), and for the one with largest total number of transactions during the full six-months window of case $A$ (USDC-WETH/500).}
\label{fig:max-tvl-txns}
\end{figure}

To gain a brief insight into the most liquid and active venues, we consider the pools extracted for case $A$ and plot in Fig. \ref{Barplots-info-tvl} the $10$ pools with highest proxyTVL at the end of June 2022, and in Fig. \ref{Barplots-info-txn} the $10$ pools with highest total number of transactions over the six months of relevance. For the first pool in the ranking of both measures, we plot the related evolution of liquidity and daily number of transactions in Fig. \ref{fig:max-tvl-txns}. As a convention, we refer to pools with the format \say{SYMBOL1-SYMBOL2/feeTier}, where we use the trading symbols of the two tokens exchanged by the pool.
Stablecoins, wrapped Ether (WETH) and wrapped Bitcoin (WBTC) dominate the landscape of tokens swapped in the most liquid and active venues, which is expected since they are the oldest, most established, or safest cryptocurrencies that agents can trade and develop strategies onto.  Then, it is interesting to observe how the DAI-USDC/100 pool is much younger than the USDC-WETH/500 one, but quickly gained strong liquidity levels due to its tokens being both stablecoins.

\paragraph{Filtering of pools by interconnectedness.} Considering the full list of data sets of liquidity provision and consumption actions to pursue investigations of more than $100$ pools becomes quickly computationally expensive. Thus, we further subset our pools of interest by requiring minimum levels of interconnection between them. This is also \dhm{beneficial in assuring a focus on the joint dynamics} that characterise the Uniswap ecosystem as a whole.

\begin{figure}[t]
    \centering
    \includegraphics[width=0.8\linewidth]{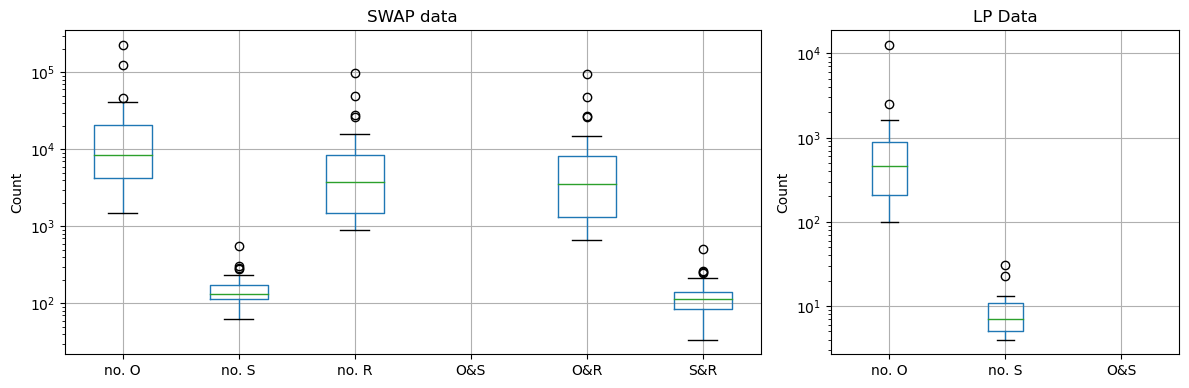}
    \caption{The intersection of origins and senders is always zero since the former are wallets of users and the latter smart contracts. Recipients can instead be both, hinting to more complex patterns in the execution of transactions. To be precise, here we are  specifically considering the sub-universe of pools relevant for case $A$ at the end of all our refinements.}
    \label{fig:OSR}
\end{figure}

For each one of our cases $A/B1/B2/C1/C2/C3$, we build a weighted graph $G=(P,E)$. The set of nodes $P$ denotes relevant pools, and edges $(p, q) \in E$ with $p,q \in P$ have weights $w_{pq}$ that encode a measure of similarity defined below. We start by considering two possible different measures of connection between pools:
\begin{enumerate}
    \item Number of common LTs (\textit{or} LPs) active on both pools, which are identified by the entry \say{origin} in the Uniswap data.
    \item Number of common smart contracts, i.e. \say{senders} in the Uniswap data, called by origins to execute swap transactions (\textit{or} to execute liquidity provision operations).
\end{enumerate}
We separate between a focus on liquidity consumption or provision in the above measures, since the two dynamics differ substantially, and one might prefer to enhance the sub-universe under consideration to be more representative of one or the other. Of course, the intersection or union of the results can be then used to pursue broader analyses.
To clarify some Uniswap terminology,  every \say{swap action} is initiated by an \textit{origin O}, then it calls a smart contract referred to as \textit{sender S}, and ends to the \textit{recipient R}. In liquidity provision, only the origin and sender of operations are relevant.
Figure \ref{fig:OSR} shows the distribution of number of origins, senders and recipients in each pool for both LT and LP data over the time window of case $A$. We also show the distribution of the intersection between origins, senders and recipients' addresses.

\begin{figure}[t]
    \centering
    \begin{subfigure}{.5\textwidth}
      \centering
      \includegraphics[width=.99\linewidth]{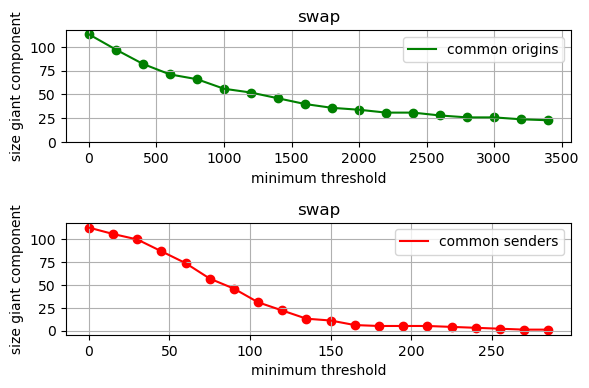}
      \caption{LT data.}
      \label{fig:sub1}
    \end{subfigure}%
    \begin{subfigure}{.5\textwidth}
      \centering
      \includegraphics[width=.99\linewidth]{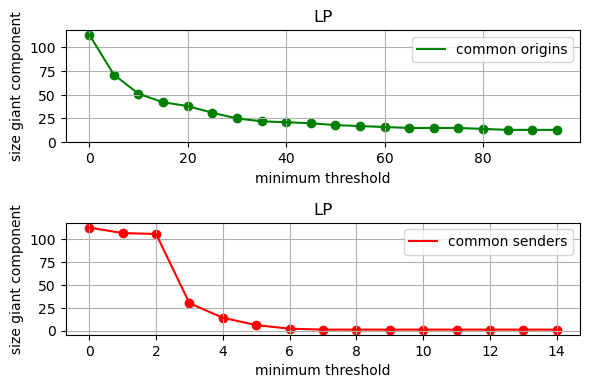}
      \caption{LP data.}
      \label{fig:sub2}
    \end{subfigure}
\caption{Evolution of the size of the giant component for graphs of pools in case $A$, when varying the threshold of common origins and senders for (a) swap transactions, (b) liquidity provision operations.} 
\label{fig:largest-connected-component}
\end{figure}

We now proceed to studying the relationship between the size of each graph's giant component and a minimum threshold on the value of the measure used to create the link between each pair of  pools. After fixing a threshold, we consider the pools in the related giant component as our relevant interconnected sub-universe. 
Figure \ref{fig:largest-connected-component} shows the variation in size of the giant component for case $A$, when modifying the minimum number of common origins or senders for LT and LP data.
We aim at considering the tails of the distributions for each case (i.e. time interval), which amounts to $\sim 20-30$ pools in each instance, to retain the most significant connections and possible dynamics of the ecosystem. 
For case $A$, this results in the choice of thresholds $2,000$ and $100$ for minimum common origins and common senders respectively, on the LT data. Similarly, we choose thresholds $30$ and $3$ for minimum common origins and common senders respectively, on the LP data.
Finally, we consider the intersection of survival pools for the two graphs generated by LT data,  and find $27$ common pools (out of the $34$ and $36$ pools, respectively in each graph). For LP data, we find a number of $19$ final relevant pools (from the intersection of $25$ and $30$ pools). The full pipeline is repeated for cases $B1/B2/C1/C2/C3$.

\begin{figure}[h]
    \centering
    \includegraphics[width=0.8\linewidth]{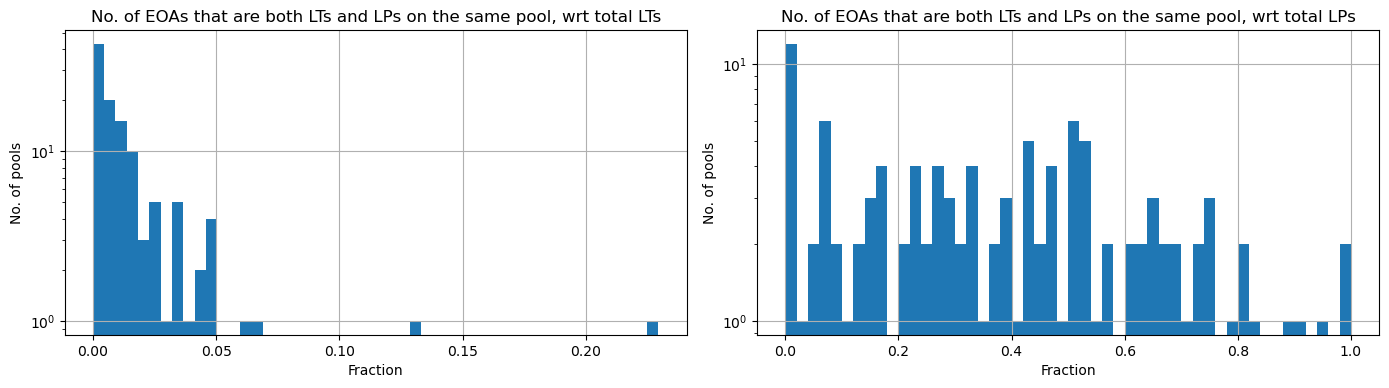}
    \caption{Distribution of \dm{origins}
    that act both as LTs and LPs on the same pool.}
    \label{fig:commonLT-LP}
\end{figure}

For the interested reader, Fig. \ref{fig:commonLT-LP} depicts the distribution of origins, also called Externally Owned Accounts (EOAs), that are both LTs and LPs on each pool for case $A$ before filtering by interconnectedness. This quantity is shown both as a ratio of the total number of LTs and of LPs on each pool.
We witness an extreme case for pool \say{WETH-sETH2/3000}, for which more than $20\%$ of the total amount of LTs are also LPs. However, the number of LTs that also act as liquidity providers is a negligible minority. If we further count the total number of LTs, LPs and LTs acting also as LPs regardless of the pool, we find  $479,161$ and $23,952$ and $13,640$ such market participants, respectively.  Approximately half of LPs also act as LTs, but LTs acting also as LPs are still a small minority.

\paragraph{Final enhancement on pools for liquidity consumption analysis.} Ideally, we should also consider the flow of funds across pools and find the related most interconnected graph. However, this is intractable if using only Uniswap data and not the full list of Ethereum blockchain transactions. Indeed, LTs are active across different DeFi protocols and can easily move liquidity from one venue to another and back.
We propose an approximation to the problem taking advantage of the fact that each trader's transaction can include more actions, which happen \say{instantaneously but in order} when the full transaction is validated.
Thus, if a LT executes two swaps of the form $X \rightarrow Y, Y \rightarrow Z$ for tokens $X,Y,Z$ in one same transaction, then we interpret $Y$ as a \textit{bridge} between the action of selling $X$ to buy $Z$. We view this as an indication of the flow of (smart) money between pools and of possible arbitrage opportunities, relevant to the LT sub-universe. 
In summary, we consider the following steps:
\begin{enumerate}
    \item Merge all LT data before the  interconnectedness analyses, e.g. data for the $113$ pools of case $A$.
    \item Keep all the transactions for which there are at least two inner actions, i.e. same \say{transaction id} but different \say{logIndex} in Uniswap terminology.
    \item For each resulting transaction:
    \begin{enumerate}
        \item For each token that appears in the transaction actions, keep a \textit{flow list} of related buying ($-1$) or selling ($+1$) trades in all the related pools by looking at the sign of the amount swapped by the pool.
        \item For each token, consider its flow list and find all the occasions when a $-1$ is immediately followed by a $+1$ (i.e. the token was first bought in a pool and then sold in another pool, acting as one of our bridges).
        \item Save this occurrence of a flow between pools as a \textit{bridge transaction},
    \end{enumerate}
\end{enumerate}
where we are approximating only jumps of length one. As an example, for a flow list of the form $[-1, +1, +1]$, we only consider the flow as one from the first pool to the second one. For more specific analyses, one could consider the specific amounts traded and check the relative proportions exchanged from the first pool to the second and third ones, but this is outside the scope of our current investigation.

\begin{figure}[h]
    \begin{subfigure}{.6\textwidth}
      \centering
      \includegraphics[width=.99\linewidth]{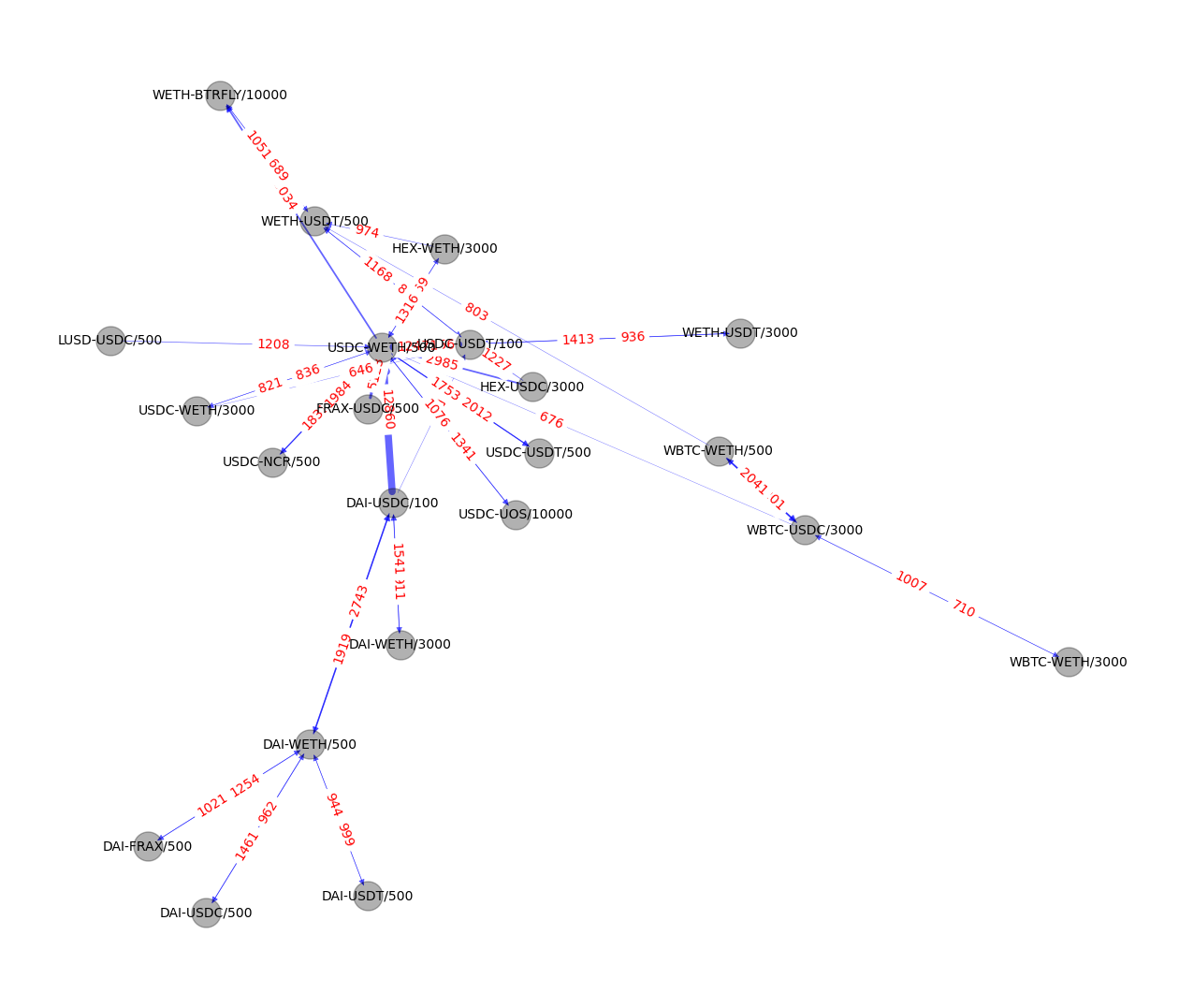}
      \caption{\mc{The resulting giant component, with edge weights reported in both directions.} }
      \label{fig:graph-bridges1}
    \end{subfigure}%
    \begin{subfigure}{.4\textwidth}
      \centering
      \includegraphics[width=.99\linewidth]{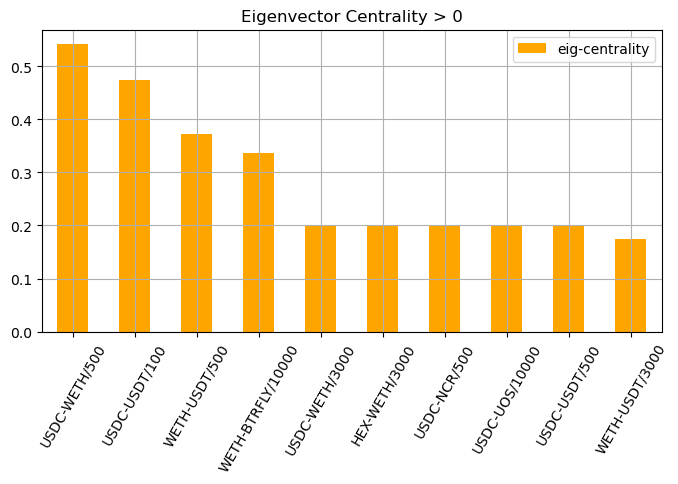}
      \caption{Pools with highest eigenvector centralities.}
      \label{fig:graph-bridges2}
    \end{subfigure}
    \caption{Results from our bridges investigation for case $A$, \dm{which covers the six-months window from January to June 2022. If a LT executes two swaps $X \rightarrow Y, Y \rightarrow Z$ one after the other (for tokens $X,Y,Z$), then we interpret $Y$ as a bridge between the action of selling $X$ to buy $Z$. We save all pairs of pools for which there is a common token that acts as a bridge, with the related number of occurrences of bridge transactions. Then, we create a directed graph where nodes are pools and edges are built for each pair of pools that have at least $800$ bridge transactions between them.}
    }
    \label{fig:graph-bridges}
\end{figure}

We extract all bridge transactions between pools and create a directed graph for each one of our temporal cases. Nodes are pools as usual, and edges are built for each pair of pools that have at least some number of bridge transactions between them. Of course, each pair of pools can have up to two edges between them according to the direction of related bridge transactions. Then, we keep the largest connected component from the undirected version of the graph and add the resultant set of nodes to the LTs pools saved from the previous interconnectedness analyses.
For case $A$, we require at least $800$ bridge transactions between two pools to create the related edges. The resultant giant component (see Fig. \ref{fig:graph-bridges1} for a visualisation) has $22$ nodes, seven of which were not included in our LT set of pools from the previous analyses and are thus added. Figure \ref{fig:graph-bridges2} further highlights the nodes with highest eigenvector centrality in the graph, where we can especially notice how several pools of WETH against a stablecoin are proposed. This is intuitively sensible, since LTs can take advantage of routing to complete specific re-balancing of tokens via more liquid and favourable pools, which tend to have stablecoins, WETH and WBTC as their tokens, as shown in the earlier analyses.

The final list of $34$ pools that we propose to consider for LTs analyses is:
DAI-WETH/3000, CEL-WETH/3000, USDC-UOS/10000, DAI-USDC/100, SPELL-WETH/3000, WETH-CRV/10000, USDC-USDT/500, DAI-FRAX/500, WETH-BTRFLY/10000, GALA-WETH/3000, WETH-USDT/3000, WBTC-USDC/3000, DAI-USDT/500, UNI-WETH/3000, WETH-ENS/3000, DAI-USDC/500, WBTC-WETH/500, MATIC-WETH/3000, DAI-WETH/500, WETH-USDT/500, USDC-WETH/500, LINK-WETH/3000, WBTC-WETH/3000, FXS-WETH/10000, FRAX-USDC/500, USDC-WETH/3000, USDC-WETH/10000, LUSD-USDC/500, HEX-USDC/3000, USDC-NCR/500, SHIB-WETH/3000, DYDX-WETH/3000, USDC-USDT/100, HEX-WETH/3000.

Regarding LP pools, we have instead the following $19$ pools: WETH-CRV/10000, MKR-WETH/3000, WETH-USDT/3000, WBTC-USDC/3000, UNI-WETH/3000, WETH-ENS/3000, WBTC-WETH/500, MATIC-WETH/3000, DAI-WETH/500, WETH-USDT/500, USDC-WETH/500, LINK-WETH/3000, WBTC-WETH/3000, USDC-WETH/3000, SHIB-WETH/3000, WBTC-USDT/3000, USDC-USDT/100, USDC-USDT/500, SHIB-WETH/10000.

The final results for cases $B1/B2/C1/C2/C3$ are then listed in Appendix \ref{appendix:A}, for the benefit of the reader that can use these sub-universes of pools as starting point for their own investigations on Uniswap v3. In our next steps, we specifically focus on the pools extracted for longest cases $A/B1/B2$.

\section{Structural Investigation of the Uniswap v3 Ecosystem}
\label{sec:cluster}


\subsection{Clustering of Liquidity Takers}

\subsubsection{Overview and pre-processing}

The DeFi ecosystem has grown increasingly complex in the recent years. The first step to shed more light on its intrinsic features and dynamics is to better understand its own \mc{components, which is what motivates the following empirical investigation of LTs trading behaviour on Uniswap v3. This is a non-trivial task, for a number of reasons. First of all,   agents can easily generate numerous crypto wallets, and hence in some sense, \say{multiply} their identities to hide or obfuscate} their full behaviour. Their actions are then generally spread over a broad set of possible pools, vary significantly in size both 
within
and across different types of pools, and also happen with evolving frequencies over time. Applying the usual initial clustering methodologies would indeed be difficult (i.e. defining a set of features that characterise pools to then perform dimensionality reduction, and finally compute similarity measures), due to the complexity of the ecosystem.
Thus, we propose a novel method to express and cluster structural trading equivalence of agents on multiple environments by leveraging both network analysis and NLP techniques. A sample of \mc{possibly external features is then used to judge and characterise} the groups unravelled and extract insights on the main types of agents present in the ecosystem.

\begin{figure}[h]
    \centering
    \includegraphics[width=0.7\linewidth]{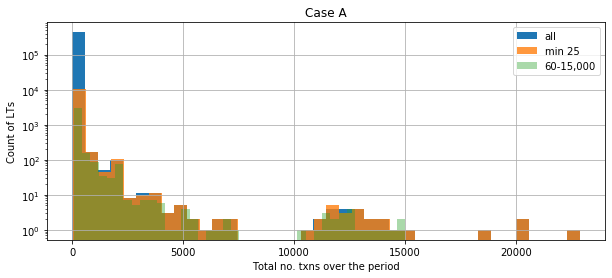}
    \caption{Distribution of total number of transactions (txns) performed by LTs during case $A$. We show the full distribution, the result after requiring a minimum of at least $25$ transactions, and the distribution after applying thresholds of minimum $60$ and maximum $15,000$ transactions. The latter scenario results in our final set for case $A$, which comprises $3,415$ LTs. A small cluster of LTs much more active than others is already discernible.}
    \label{fig:LTs-txns}
\end{figure}

We focus on the LT data for our three longest \mc{periods} $A/B1/B2$, which we defined and described in the previous section. 
For each case, we first look at the distribution of the total number of transactions performed by the different LTs over each full time window.
\dhm{As an example, the distribution for case $A$ is represented by the blue bars in Fig. \ref{fig:LTs-txns}.}
We then require a minimum number of transactions completed by each LT, since \mc{considering} only a very small sample of trades per agent does not provide meaningful structural information on their behaviour. Thus, we \dhm{impose a lower bound of at least an average of $10$ transactions per month that each LT must have completed.
On the other hand, we manually define maximum thresholds to remove only extreme singular outliers from each distribution for computational purposes}.The initial total distribution for case $A$ is shown in Fig. \ref{fig:LTs-txns}, where we also highlight how it changes when requiring a minimum number of transactions equal to $25$ \dhm{(orange bars)}, and when we require our final minimum and maximum thresholds of $60$ and $15,000$ total number of transactions, respectively \dhm{(green bars)}. For cases $B1/B2$, we require the range $30$ to $5,000$ transactions for the former, and $30$ to $11,000$ transactions for the latter. Overall, we find a number of LTs approximately between $3,500$ and $5,000$ for all our periods $A/B1/B2$. 
\mc{This altogether defines the final sets of LTs along with their transactions. Next, we proceed to computing their embeddings, which are subsequently used for the final clustering stage.}

\subsubsection{Methodology}


\paragraph{NLP background and \textit{graph2vec}.} The field of Natural Language Processing (NLP) studies \mc{the development of algorithms for processing, analysing, and extracting} meaningful insights from large amounts of natural language data. Examples \mc{of its myriad applications include} sentiment analysis of news articles, text summary generation, topic extraction and speech recognition. One of the turning points in NLP was the development of the \textit{word2vec} word embedding technique \cite{mikolov2013efficient}, which considers sentences as directed subgraphs with nodes as words, and uses a shallow two-layer neural network to map each word to a unique vector. The learned word representations capture meaningful syntactic and semantic regularities, and if pairs of words share a particular relation then they are related by the same constant offset in the embedding space. As an example, the authors observe that the singular/plural relation is captured, e.g. $x_{apple}-x_{apples} \approx x_{car} - x_{cars}$, where we denote the vector for word $i$ as $x_i$.
Words sharing a common context in the corpus of sentences also lie closer to each other, and therefore, relationships such as $x_{king}-x_{man}+x_{woman} \approx x_{queen}$ are satisfied with the analogies indeed predicted by the model.

Taking inspiration from this idea of preserving knowledge of the context window of a word in its embedding, the \textit{node2vec} algorithm \cite{grover2016node2vec} learns a mapping of nodes in a graph to a low-dimensional space of features by maximising the likelihood of preserving network neighbourhoods of nodes. The optimisation problem is given by 
\begin{equation}
    \max_{f} \sum_{s \in S} \log Pr(N_{L}(s)|f(s)), 
    \label{node2vec}
\end{equation}
where $G=(S,T)$ is a graph with nodes $S$ and edges $T$, $f$ is the mapping function for nodes to $n$-dimensional vectors that we aim to learn, and $N_L(s) \subset S$ is the network neighbourhood of node $s$ generated with sampling strategy $L$. The latter is designed by the authors of \textit{node2vec} as a biased random walk procedure, which can be tuned to either focus on sampling a broader set of immediate neighbours, or a sequence of deeper nodes at increasing distances.
Then, Problem \eqref{node2vec} is solved for $f$ by simulating several random walks from each node and using stochastic gradient descent (SGD) and backpropagation.

By taking a further step towards general language representations, \cite{doc2vec} proposes the unsupervised algorithm \textit{Paragraph Vector} (also known as \textit{doc2vec}), which learns continuous fixed-length vector embeddings from variable-length pieces of text, i.e. sentences, paragraphs and documents. The vector representation is trained to predict the next word of a paragraph from a sample of the previous couple of sentences. Both word vectors and paragraph vectors need to be trained, which is again performed via SGD and backpropagation. 

As \textit{doc2vec} extends \textit{word2vec}, \textit{graph2vec} \cite{graph2vec} is a neural embedding framework that aims to learn data-driven distributed representations of an ensemble of arbitrary sized graphs.
The authors propose to view an entire graph as a document, and to consider the rooted subgraphs around every node in the graph as words that compose the document, in order to finally apply \textit{doc2vec}. This approach is able to consider non-linear substructures and has thus the advantage to preserve and capture structural equivalences.
One necessary requirement to pursue this analogy is for nodes to have labels, since differently labelled nodes can be then considered as different words. These labels can be decided by the user, or can be simply initiated with the degree of each node. Thus, \textit{doc2vec} considers a set of graphs $\mathcal{G}=\{G_1, G_2...\}$, where the nodes $S$ of each graph $G=(S,T, \lambda)$ can be labelled via the mapping function $\lambda: S\rightarrow \mathcal{L}$ to the alphabet $\mathcal{L}$.
The algorithm begins by randomly initialising the embeddings for all graphs in the set $\mathcal{G}$, then proceeds with extracting rooted subgraphs around every node in each one of the graphs, and finally iteratively refines the corresponding graph embedding in several epochs via SGD and backpropagation, in the spirit of \textit{doc2vec}.
The rooted subgraphs act as the context words, which are used to train the paragraph (i.e. graph) vector representations. Subgraphs are extracted following the Weisfeiler-Lehman (WL) relabeling process \cite{WL-relabel}. The intuition is that, for each node in a graph, all its (breadth-first) neighbours are extracted up to some depth $d$. Labels are then propagated from the furthest nodes to the root one, and concatenated at each step. In this way, a unique identifier for each node is identified from its \say{context} and the full set can be used to train an embedding for the graph.
The optimisation problem thus becomes
\begin{equation}
    \max_{f'} \sum_{G \in \mathcal{G}} \sum_{s \in S} \log Pr(g_{WL}^{d}(s)|f'(G)), 
    \label{graph2vec}
\end{equation}
where the aim is to maximise the probability of the WL subgraphs given the current vector representation of the graph. Here, $f'$ is a mapping function of graphs to $n$-dimensional representations, and $g_{WL}^{d}$ are WL subgraphs with depth $d$.


\paragraph{A modification of \textit{graph2vec} for LTs embedding.} For each one of our cases $A/B1/B2$, we consider all the related LTs and their full set of transactions on the sub-universe of LTs' pools of relevance. 
We then introduce the concept of a \textit{transaction graph $G_{txn}$}, which we use to represent the behaviour of each active agent. 

\begin{definition}[Transaction graph]
A \textit{transaction graph} $G_{txn}=(S, T, W)$ is the complete \mc{weighted} graph where nodes $S$ are the swap actions that the LT under consideration has executed, and edges $(s,r) \in T$ with $s,r \in S$ are built between every pair of nodes. Each edge has a weight $w_{sr} \in W$, which \mc{encodes the amount of time $\Delta t$ (in seconds) elapsed between the two transactions} $s,r$. Each node $s \in S$ has a label $l_s$ from the alphabet $\mathcal{L}$ 
, which uniquely identifies the pool that the swap was executed into. Importantly, $\mathcal{L}$ is shared among the full set of LTs and related transaction graphs. 
\end{definition} 

Labels in the alphabet $\mathcal{L}$ differentiate between swaps executed on different pools, i.e. pools with unique combination of tokens exchanged and feeTier implemented. This implies that the algorithm receives as input only \dm{general identifiers of pools. Thus, we}
we can consider intuitive differences (e.g. expected volatility of the exchange rate on pools of stablecoins versus on pools of more exotic tokens) only afterwards, when assessing and investigating the meaningfulness \mc{and interpretability of the extracted} clusters.

We now have a set of graphs representing LTs, and our aim is to find a $n$-dimensional vector representation of each one of its elements. We cannot plainly apply the \textit{graph2vec} algorithm, since the concept of neighbours of a node is irrelevant in a complete graph. 
Thus, we modify its mechanism to take advantage of the weight that the different links between nodes have, while maintaining the overall intuition. For each node $s \in S$ of a graph $G_{txn}$, we sample a set of neighbours $N_{txn}(s)$ by generating random numbers from a uniform distribution between $[0,1]$ and comparing them to the \textit{cut-value} of the edges between the node and possible neighbours. If the value is below the cut-value, then the link is kept and the associated node added to $N_{txn}(s)$. In this way, the probability of an edge to be chosen is inversely proportional to its weight $\Delta t$, and the sub-structures kept represent clustered activity in time. 

\begin{definition}[Cut-value]
The \textit{cut-value} $C(w_{sr})$ of an edge $(s,r) \in T$ with weight $w_{sr} \in W$ in graph $G_{txn}=(S, T, W)$ is computed as
\begin{equation}
\begin{split}
    C(w_{sr}) &= \frac{H(f^{scal}(w_{sr}))}{H(f^{scal}(\min W))},  \\
    \mbox{with} \quad   \; H(w_{sr}) &= \sqrt{\frac{2}{\pi}}\exp{\frac{-w_{sr}^2}{2}}, w_{sr} \geq 0, \\
    f^{scal}(w_{sr}) &= \frac{w_{sr}-\min W}{(\max W)/|S|}, \end{split}
\label{cut-value}
\end{equation}
where we are using a half-norm that is shifted and scaled to adapt to each LT's extreme features, i.e. $\min W$ and $\max W$. The final cut-value is also normalised to impose a value of $C(\min W)=1$, meaning that the shortest link(s) in the graph is chosen with probability $1$ (of course, only if it is involved in the current node under consideration).
\end{definition}

After having generated the set of $N_{txn}(s),  \forall s \in S$, we perform  WL relabeling and proceed as in the vanilla version of the \textit{graph2vec} algorithm. We set all the hyperparameters to their default values, i.e. number of workers = 4, number of epochs = 10, minimal structural feature count = 5, initial learning rate = 0.025, and down sampling rate of features = 0.0001. The only exception is the number of WL iterations, which in our case must be set to $1$ instead of $2$.
The result is an embedding for each graph in our set of transaction graphs, which becomes a set that we can subsequently cluster via the popular  k-means++ methodology. 
Importantly, we want to underline that our embeddings and clusters do not depend on the real magnitudes of weights $\Delta t$, since the sampling is adjusted on that. \mc{In addition, they also have no notion of the amount of USD traded, thus being agnostic to the transaction value.} 
As a final note, we refer the reader to \cite{dual-graph2vec} for a version of \textit{graph2vec} that uses edge labels. However, the algorithm creates the dual version of the graph and would not be effective in our case, \mc{thus providing ground for our proposed extension.}  

\begin{figure}[h]
\centering
\begin{subfigure}{.4\textwidth}
  \centering  \includegraphics[width=.9\linewidth]{new/explanatory.png}
  \caption{Transaction graph $G_{txn}=(S,T,W)$.}
  \label{fig:diagram}
\end{subfigure}%
\begin{subfigure}{.6\textwidth}
  \centering
  \includegraphics[width=.8\linewidth]{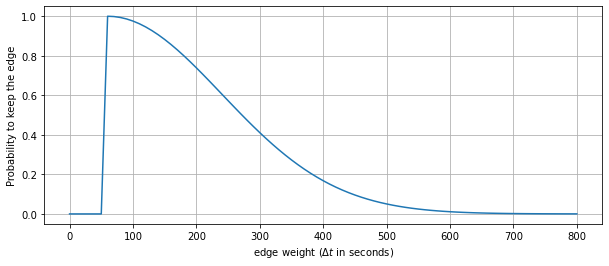}
  \caption{Cut-value $C(w)$.}
  \label{fig:halfnorm}
\end{subfigure}
\caption{For our illustrative example, we show in (a) a simplified representation of the LT's transaction graph, and in (b) the cut-value that defines probabilities of keeping edges as neighbours.} 
\label{fig:explanatory}
\end{figure}

\paragraph{\mc{An illustrative example}.} To \dhm{clarify}
our approach, we describe a simple example. Consider an agent that \dhm{executes} $20$ transactions. She \dhm{executes} the first $10$ transactions shortly clustered in time, waiting only $60$ seconds one after the other. Then, she waits $42$ minutes to action on the final $10$ transactions with, again, a frequency of one minute. Her behaviour is plotted in the transaction graph $G_{txn}$ of Fig. \ref{fig:diagram}, where we assume for simplicity that each transaction is performed on the same pool and thus colour-code nodes all the same. We also number nodes to show the order in which the related transactions are executed. We do not draw all the edges of this complete graph for \dhm{clarity}
\mc{and ease of visualisation}, but hint with the green dashed lines that indeed there are more connections to be remembered.  
In this example, the minimum time between transactions is $60$ seconds (light blue edges) and the maximum one is one hour, i.e. $3,600$ seconds (light grey edge). Some intermediate times are depicted as edges with the same colour for the same weight. The resultant cut-value function $C(w)$ that defines our sampling probabilities to choose edges is shown in Fig. \ref{fig:halfnorm}. As intuitively desired, we aim at always keeping the shortest edges and indeed these have probability $1$. Then, we also aim  to keep the most clustered \say{communities}, and indeed, we observe from the plot that transactions five minutes away are still chosen with $40\%$ probability, but longer times are very easily dropped.

\subsubsection{\dhm{Analysis} of results}

For each case $A/B1/B2$, we study the structural equivalence of LTs' trading activity by clustering the representations generated via our modified \textit{graph2vec} algorithm.
Focusing first on case $A$, we compute embeddings for dimensions $n \in \{8, 16, 32, 64\}$, and confirm with Principal Component Analysis (PCA) that the proportions of data's variance captured by different dimensions are well-distributed. For each $n$-dimensional set of vectors, we then group LTs by performing a series of k-means++ clusterings with different number of desired groups. 
\dhm{We compute the inertia of each partition found, where inertia is defined as the sum of squared distances of samples to their closest cluster center. Then, we choose the optimal clustering via the \textit{elbow method}, i.e. by picking the point where the marginal
decrease in the loss function (inertia) is not worth the additional cost of creating another cluster.}
The similarity between optimal clusterings for different dimensions is then computed, in order to investigate the stability of results across representations of increasing dimensionality. We achieve this by computing the Adjusted Rank Index (ARI) \cite{ARI}, which is a measure of similarity between two data clusterings in the range $[-1, 1]$, adjusted for the chance of grouping of elements. We find ARIs for clusterings on $8$-vs-$\{16, 32, 64\}$ dimensional data around $0.75$, while clusterings on $16$-vs-$32$, $16$-vs-$64$ and $32$-vs-$64$ dimensional data reach approximately the value of $0.90$. Therefore, we conclude that there is a high stability of results when our data are embedded at least in $16$ dimensions, and use the related $16$-dimensional vector representations for our final analyses.  The related optimal number of clusters of LTs for case $A$ is seven.
Similar results arise for cases $B1/B2$ too, and the related optimal numbers of clusters of LTs are six and seven, respectively.

Each extracted clustering is based on the structural similarity of LTs' trading behaviour. To judge the goodness of our modified algorithm and assess the results, we investigate whether there are specific features or trends that are highly  representative of only some of the groups.
Thus, we proceed to computing a set of \dhm{summary statistics} for each LT, and calculate the average of these results over the LTs belonging to each different group. 
The features that we consider are:
\begin{itemize}
    \item average and median USD traded,
    \item average and median time $\Delta t$ in seconds between transactions,
    \item proportion of transactions done in \say{SS}, \say{EXOTIC} or \say{ECOSYS} pools, and related entropy,
    \item proportion of transactions done in pools with a specific feeTier, and related entropy,
    \item proportions of trades on days when the SP LargeCap Crypto Index\footnote{\url{https://www.spglobal.com/spdji/en/indices/digital-assets/sp-cryptocurrency-largecap-index/\#overview}} \mc{increased or decreased in value, or when} the market was closed, due to weekends and bank holidays.
\end{itemize}
The distinction between \say{SS}, \say{EXOTIC} or \say{ECOSYS} pools is inspired by the classification in \cite{behaviour-LP}, where the authors introduce a notion of normal pools, stable pools and exotic pools. For them, stable pools exchange tokens that are both stablecoins. Normal pools trade instead tokens that are both recognised in the crypto ecosystem, while exotic pools deal with at least one token that is extremely volatile in price (e.g. YAM, MOON and KIMCHI).
We slightly divert from this classification and define \say{SS} pools as pools whose tokens are both stablecoins, \say{ECOSYS} pools as pools that exchange only tokens that are either stablecoins or pegged to the most established BTC and ETH coins, and \say{EXOTIC} pools as the remaining ones.
ECOSYS pools can be seen as the venues carrying the \say{safest} opportunity for profit for a novice 
crypto investor, since they trade volatile tokens though directly related to the most established blockchains that are the true foundations of the whole DeFi environment.

\begin{figure}[t]
\centering
\begin{subfigure}[c]{.75\textwidth}
  \centering  \includegraphics[width=.99\linewidth, valign=t]{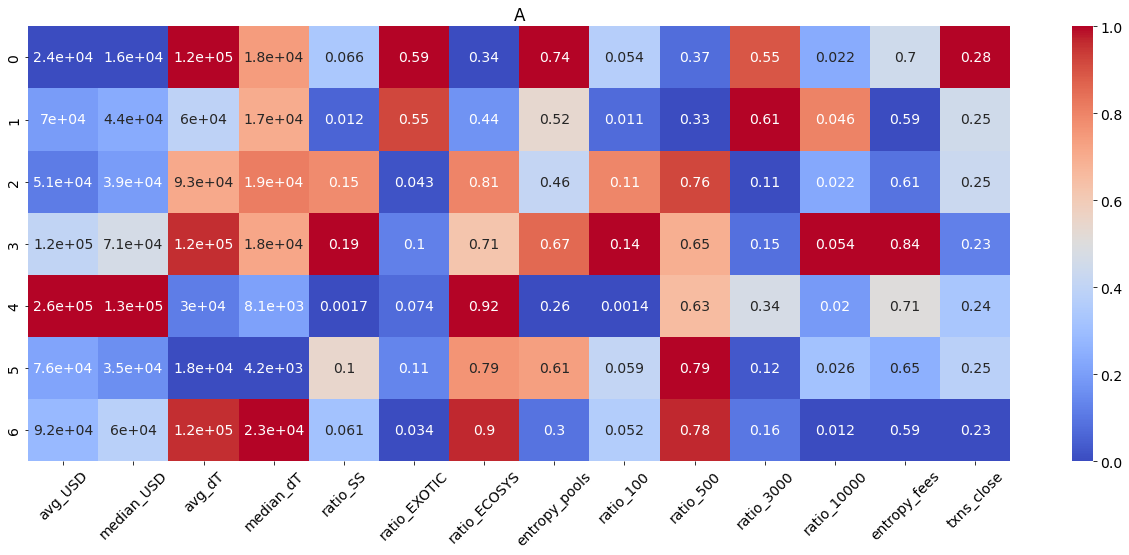}
  \caption{Average features for LTs in case $A$. 
  }
  \label{fig:heatmapA}
\end{subfigure} \\
\vspace{0.2cm}
\begin{subfigure}[c]{.5\textwidth}
  \centering
  \includegraphics[width=.99\linewidth, valign=t]{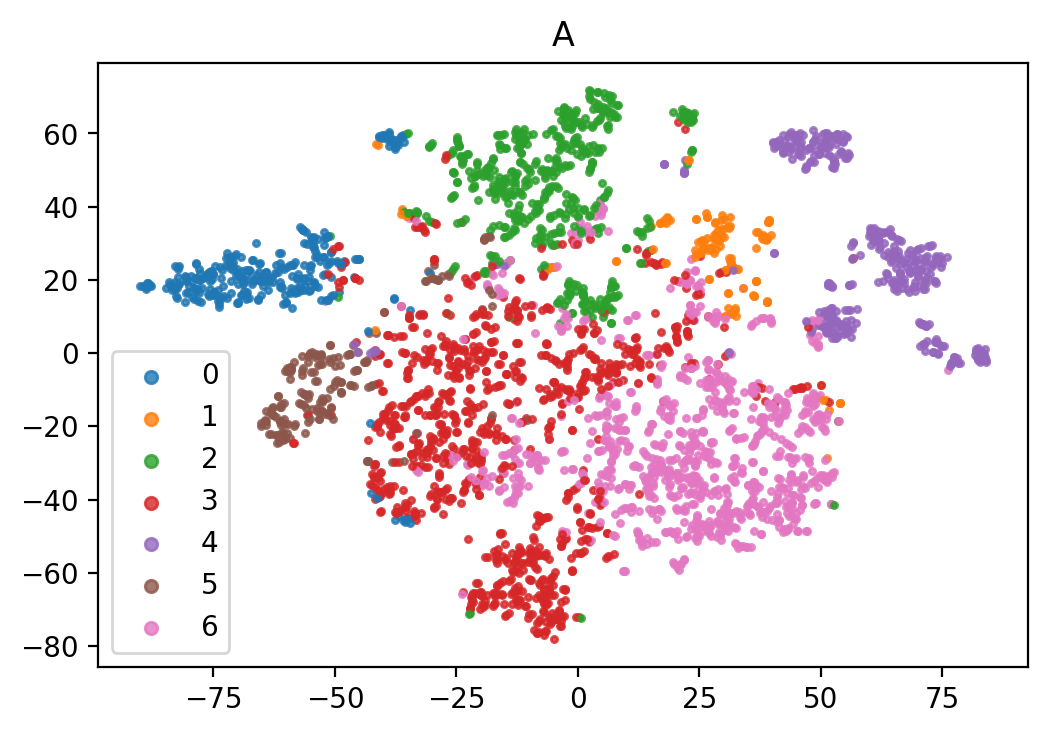}
  \caption{t-SNE embedding visualisation of case $A$.
  }
  \label{fig:tsne}
\end{subfigure}
\caption{Clustering of LTs for case $A$, i.e. over the six-months time window between January and June 2022. In (a), each row represents one of the recovered clusters and columns are the different features computed to characterise species of LTs. The color-code employed applies to each column separately to be able to quickly identify the related smallest and biggest values in magnitude, and judge the general distribution. It is essential to always check the magnitudes of cells per se too, due to highly variable variance between columns. In (b), the t-SNE plot of embeddings of LTs is reported with perplexity = $15$ and points are color-coded according to their cluster of membership.}
\label{fig:clusteringA}
\end{figure}

The average magnitude of features computed over the LTs belonging to each different cluster for case $A$, i.e. over Jan-June 2022, is reported in Fig. \ref{fig:heatmapA}. We focus on the groups found specifically for this period because it is the longest one and thus, it provides us the most general results and insights.
Cases $B1/B2$ will be later described too, in order to assess the overall stability of recovered \textit{species} of LTs and highlight any specific variations due to different sub-periods in time and related pools of relevance considered.
The seven clusters of LTs found have sizes of $304/142/512/978/379/186/914$ 
agents respectively, which means that we are able to find a well-balanced
distribution of cluster sizes 
without any \mc{dominant clusters in terms of size}. Thanks to the heatmap in Fig. \ref{fig:heatmapA}, we also easily confirm that our methodology is able to extract different groups of LTs that have significant variation of behaviour with respect to the outer features defined. However, a few columns had to be dropped due to non-significance of their results. Importantly, we also recall that inner biases on ratios are present (e.g. when considering that our sub-universe does not have a uniform distribution of numbers of pools with specific feeTier), and thus we can expect more/less transactions of some type on average.
For visualisation purposes, we also embed the $16$-dimensional representations of LTs into a $2$-dimensional view via t-SNE, and plot them with perplexity = $15$ in Fig. \ref{fig:tsne}. LTs are colour-coded according to the cluster they belong to, and we indeed observe that different groups lie on different parts of the plane.

Focusing on Fig. \ref{fig:heatmapA}, one can \mc{draw the following high-level remarks}.

\begin{itemize}
    \item
\textbf{Groups 0 and 1} have a strong focus on trading exotic cryptocurrencies. The former set of LTs mainly uses feeTier $3000$ for the purpose, and shows slightly higher than average tendency to trade when the market is closed. The latter group uses significantly both the $3000$ and $10000$ feeTiers, meaning that the related LTs are willing to accept also extremely high transaction costs. This behaviour could indicate that they have high confidence on their intentions and possibly urgency.

\item 
On the other hand, \textbf{groups 2 and 3} trade stablecoins more than usual. The former cluster could point to an enhanced use of SS pools to take advantage of optimised routing, while the latter has a non-negligible proportion of trades in exotic pools with feeTier $10000$. 
\dm{Likely, group 3 isolates a set of LTs that are interested in niche exotic tokens, which are only proposed in pools against stablecoins that do not overlap. Diverting funds between two of these exotic tokens requires an exchange between the two related stablecoins too, which motivates the recovered statistics. We also witness strong usage of the feeTier $100$, which hints to traders trying to compensate the high costs suffered in pools with feeTier $10000$ by paying the lowest possible fees on the SS pools.}

\item 
 \textbf{Groups 4 and 6} are more active than average on ECOSYS pools.
The two groups differ noticeably from their opposite relative strength of USD traded and time between operations. Overall, group 6 trades less money and waits longer, mainly using pools with low feeTier $500$. 
These features can be interpreted as characteristics of cautious retail traders that invest in less risky and highly well-known crypto possibilities. And indeed, we also find that this group is one of the largest in size.
Then, group 4 also relates to ECOSYS pools. However, these users tend to trade more USD with higher frequency, and this is also the cluster with much higher than average proportion of LTs that also act as LPs ($\sim 16\%$) . Therefore, we identify here a group of more professional investors.

\item Finally, \textbf{group 5} shows a significant usage of all the three types of liquidity pools, but trades are concentrated in pools with cheap feeTier $500$. These agents trade often, and indeed show the smallest median time between transactions. These eclectic, active and thrifty LTs are probably our group of smartest investors.
\end{itemize}

Our results confirm that the proposed algorithm is able to recognise variance in the data, and allow us to extract interesting insights into the behaviour of different types of species of LTs. 
In particular, we observe how the type of pools on which LTs are active plays a primary role in the definition of their trading behaviours. This is especially interesting since no full notion of tokens and feeTier is used in the generation of the embeddings. Indeed, only a unique label per pool is provided as input to our algorithm, e.g. USDC-WETH/500 could be pool \say{P1}, USDC-WETH/3000 pool \say{P2} and FXS-WETH/10000 pool \say{P3}. Thus, these pools would be considered equally different if no structural discernible pattern was recognised by the methodology, providing some further evidence of the strength of our proposed extension to \textit{graph2vec}.



\paragraph{Stability analyses.} As already motivated, we now pursue the same analyses described above but for cases $B1/B2$. We cluster the $n$-dimensional embeddings for $n \in \{8,16,32,64\}$ and compute the ARIs between each pair of resultant sets of LT groups. 
We confirm that at least a $16$-dimensional embedding is required in order to have a stability of clusters in case $B1$, while only eight dimensions suffice for the case $B2$. For simplicity, we use the $16$-dimensional representations consistently in all cases. 
We recover six groups of LTs in case $B1$, and seven in case $B2$. In both cases, we find two clusters with same characteristics as groups $4$ and $6$ of case $A$, i.e. traders mainly active on ECOSYS pools. We also recover the eclectic traders of group $5$. 
Therefore, we observe several  stable and persistent types of LTs. Small perturbations happen instead on the groups trading on SS or EXOTIC pools, as one could expect from the mere evolution of time and external market conditions, and consequently generation of different behaviours. 
In particular, all case $A$ species, except group $1$, are also found in case $B1$. On the other hand, case $B2$ shows less intensity on group $3$, probably due to investors diversifying more during the crypto turmoils of the second quarter of 2022.
Overall, we observe general agreement on the groups and main features recovered during cases $A/B1/B2$, and we can thus rely on our species of LTs found for the longest duration case $A$ as descriptors of the ecosystem.

The above stability-related findings  are of interest in themselves, first of all, since central banks started hiking interest rates in March 2022. This consequently stopped a strong influx of liquidity into the crypto ecosystem and 
\dm{accentuated a period of significant underperformance, that could have indeed weakened the stability of results.}
On top of that, the Terra-Luna crash happened in May 2022 and it could have in theory enhanced noise and instabilities especially in the structural clustering on case $B2$.
As a very last remark, we notice that only $\sim 20\%$ addresses are present in all cases $A/B1/B2$. Therefore, we are either recovering similar behaviour but for different people, or in some cases it could be the same person simply employing a new wallet to better hide their trading behaviour.

\dhm{We have mainly focused on the liquidity consumption component of the crypto ecosystem thus far. In the next step of our investigation, we shift the focus from LTs to pools. We first aim to perform a  clustering of pools based on features built from simple statistics that consider both liquidity consumption and liquidity provision. This will allow us to assess whether the SS, ECOSYS and EXOTIC classification is beneficial for describing the crypto ecosystem or is only useful for LTs characterisation. 
}


\subsection{Clustering of pools}

\subsubsection{Motivation and Methodology}

\mc{ 
The above analyses revealed  a characterisation of the main types of LTs structural trading behaviour.
While the importance of different types of pools in the ecosystem seems to be also clear, we stress that a full understanding of liquidity pools goes beyond the mere liquidity consumption mechanism (i.e. it needs to further account for both liquidity provision and price evolution).
Thus, we now pursue an intuitive initial investigation of the similarity of pools themselves, in order to gain additional insights on the entire ecosystem. }

\begin{figure}[h]
    \centering
    \includegraphics[width=0.75\linewidth]{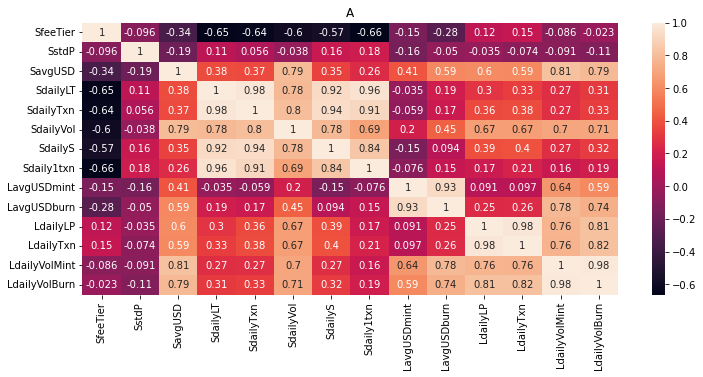}
    \caption{Spearman correlation between the computed features for pools, with the addition of feeTier, for our case $A$.}
    \label{fig:heatmap-pools}
\end{figure}

\begin{figure}[h]
\centering
\begin{subfigure}{.33\textwidth}
  \centering  \includegraphics[width=.99\linewidth]{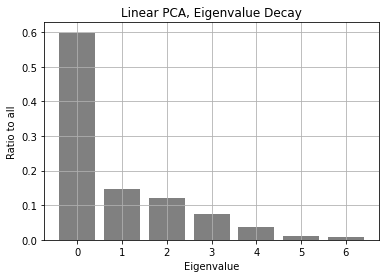}
  \caption{Linear PCA.}
\end{subfigure}%
\begin{subfigure}{.33\textwidth}
  \centering
  \includegraphics[width=.99\linewidth]{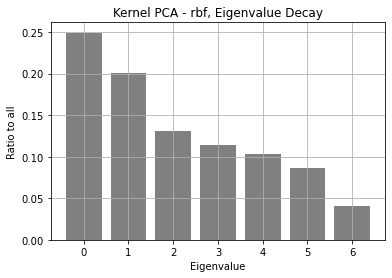}
  \caption{rbf kernel PCA.}
\end{subfigure}%
\begin{subfigure}{.33\textwidth}
  \centering
  \includegraphics[width=.99\linewidth]{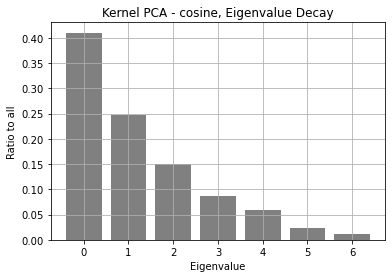}
  \caption{Cosine kernel PCA.}
\end{subfigure}
\caption{Decay of eigenvalues for the first seven out of $13$ eigenvalues for different PCA kernels.}
\label{fig:eigenv-decay}
\end{figure}

We focus on case $A$, as it covers the longest period in time. We consider the intersection of pools relevant for both LTs and LPs to properly account for both mechanisms, and find a resulting  set of $16$ pools. For each pool, we compute the following $13$ features:
\begin{itemize}
    \item average daily number of active LTs/LPs - \say{SdailyLT} and \say{LdailyLP} respectively,
    \item volatility of the execution price of the pool - \say{SstdP},
    \item average size of swap/mint/burn operations in dollars - \say{SavgUSD}, \say{LavgUSDmint} and \say{LavgUSDburn},
    \item average daily amount of dollars used in swap/mint/burn operations, i.e. volume - \say{SdailyVol}, \say{LdailyVolMint} and \say{LdailyVolBurn},
    \item average daily number of LTs/LPs transactions - \say{SdailyTxn} and \say{LdailyTxn},
    \item average daily number of different senders, i.e. smart contracts, called within swap transactions - \say{SdailyS},
    \item number of agents with only one transaction normalised by the number of days considered - \say{Sdaily1txn}. This measure is computed to gauge the tendency of external smart investors to hide their behavior by creating several different wallets on the pool.
\end{itemize}
For the above features, we create related labels for ease of reference, which start with letter \say{S} if the quantity is computed from swap operations, or letter \say{L} if the quantity is computed from liquidity provision operations.
In Fig. \ref{fig:heatmap-pools}, we show the heatmap of Spearman correlations between the above attributes plus feeTier (\say{SfeeTier}) for our pools. There are significant positive correlations, especially among features developed from LT data and LP data, respectively.
Thus, we standardise entries and employ linear PCA and kernel PCA (with both \say{rbf} and \say{cosine} kernel in the latter) to reduce the dimensionality of our data. The eigenvalue decay for all three mentioned cases is shown in Fig. \ref{fig:eigenv-decay}, where only the first seven eigenvalues are depicted for clarity of visualisation. 
The cosine kernel PCA is seen to capture more variance in fewer dimensions, and thus we embed the data by projecting  on its related first three components. The resulting 3D embedding is shown in Fig.  \ref{fig:clustering-pools-3d} from three different angles, where we color-code pools according to their feeTier. In particular, green relates to feeTier $100$, blue to $500$, orange to $3000$, and red to $10000$.

\subsubsection{Discussion}

\begin{figure}[h]
\centering
\begin{subfigure}{.33\textwidth}
  \centering  \includegraphics[width=.99\linewidth, trim=0cm 2.5cm 0cm 2.5cm,clip]{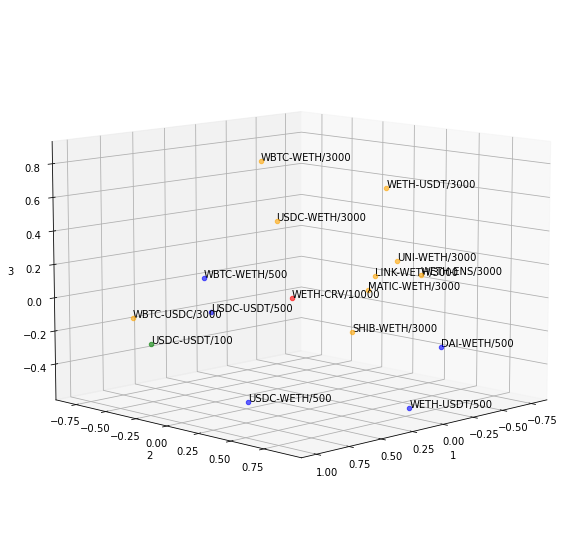}
  \caption{Angle = $45$ degrees.}
\end{subfigure}%
\begin{subfigure}{.33\textwidth}
  \centering
  \includegraphics[width=.99\linewidth, trim=0cm 2.5cm 0cm 2.5cm,clip]{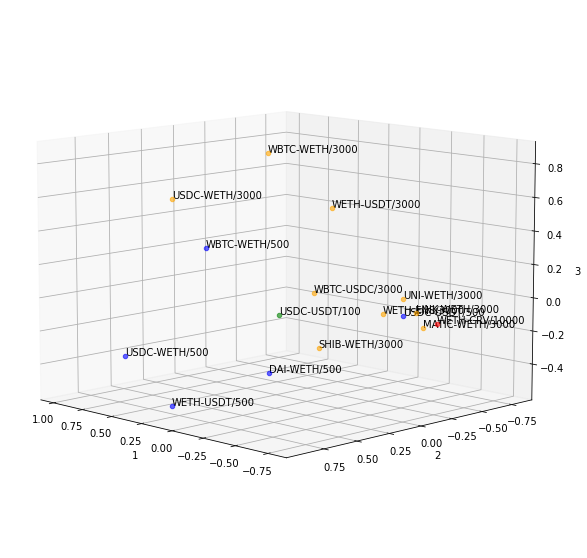}
  \caption{Angle = $135$ degrees.}
\end{subfigure}%
\begin{subfigure}{.33\textwidth}
  \centering
  \includegraphics[width=.99\linewidth, trim=0cm 2.5cm 0cm 2.5cm,clip]{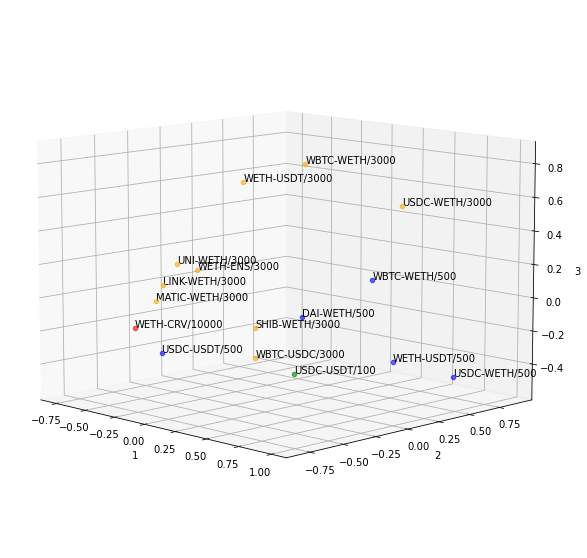}
  \caption{Angle = $315$ degrees.}
\end{subfigure}
\caption{Projection on a 3D space of the vectors encoding different features of pools, from the application of PCA with cosine kernel. Views from different angles are reported for a better judgment of the results, 
and pools are color-coded according to their feeTier (green relates to feeTier $100$, blue to $500$, orange to $3000$ and red to $10000$).
}
\label{fig:clustering-pools-3d}
\end{figure}

From the projections shown in Fig. \ref{fig:clustering-pools-3d} and initial trials of clustering, it is clear that the division between SS, ECOSYS and EXOTIC pools does not hold when considering the full set of dynamics on pools (while it is indeed suitable in connection to LTs' behaviour specifically). Similarly, we do not witness strong proximity of pools with same feeTier.
Liquidity consumption, provision and price evolution are all essential mechanisms to consider for a full description of the Uniswap ecosystem, and our intuition is that certain combinations of tokens and feeTiers are more similar and suitable for trading at different moments in time.
LPs are more incentivised to enhance liquidity on pools with strong LTs activity, low volatility of the exchange rate to avoid predictable loss, and possibly high feeTier from which they indeed mainly profit. In parallel, LTs are more interested in pools with low fees but high volatility of the price of tokens \dhm{in order to extract gains from trading opportunities}, and high liquidity to diminish the market impact of their trades. Thus, different adjustments of these mechanisms can result in the proximity or not of our projections of pools. In the next Section of our work, we propose a model to judge the health of each pool's combination of mechanisms and characterise the best venues for market participants (i.e. both LTs and LPs) to be active on.

\section{The \textit{ideal crypto law} and a \textit{cryptoness} measure  of a liquidity pool}
\label{sec:physics}

\subsection{Model: from the ideal gas law of thermodynamics to the ideal crypto law for pools}

In Physics, an ideal gas is a theoretical gas composed of many randomly moving particles with negligible volume that are not subject to interparticle interactions. On the other hand, real gases occupy space and molecules interact between themselves.
\dhm{The field of thermodynamics consists of laws that govern the behaviour of macroscopic properties of matter (e.g. of the ideal gas just mentioned above), while the aim of statistical mechanics is to derive such macroscopic properties from the laws characterising the microscopic variables and interactions of individual particles.}

\begin{figure}[t]
\centering
\begin{subfigure}{.4\textwidth}
  \centering  \includegraphics[width=.9\linewidth]{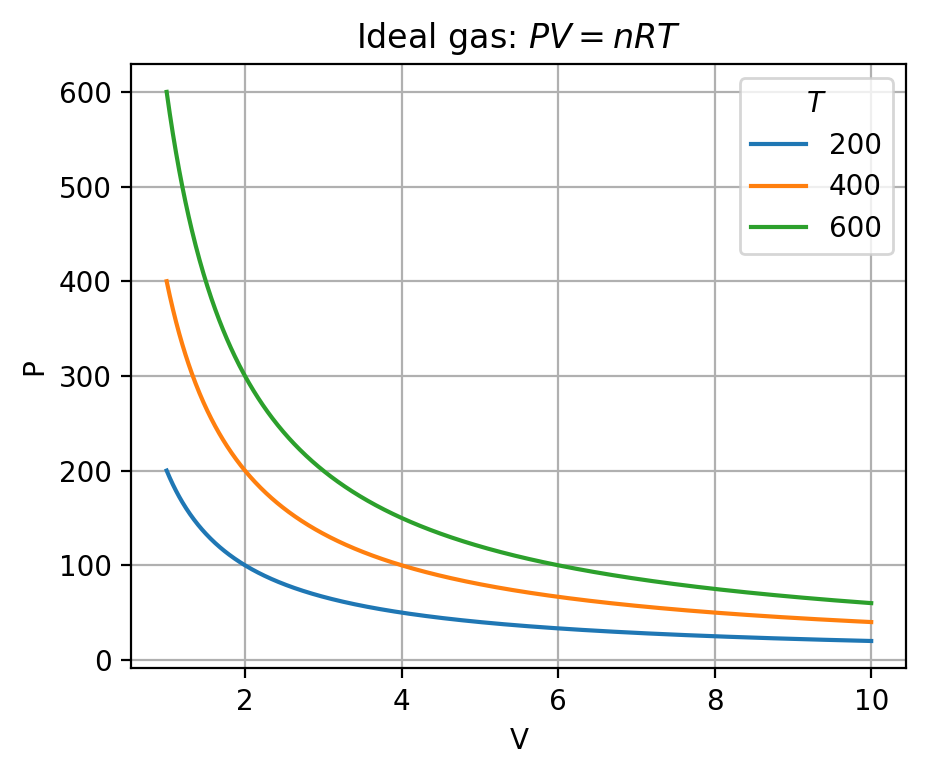}
  \caption{Sample of isotherms for an ideal gas, where we impose $nR=1$ for clarity of exposition.}
  \label{h1}
\end{subfigure}\hspace{0.1\textwidth}
\begin{subfigure}{.4\textwidth}
  \centering
  \includegraphics[width=.9\linewidth]{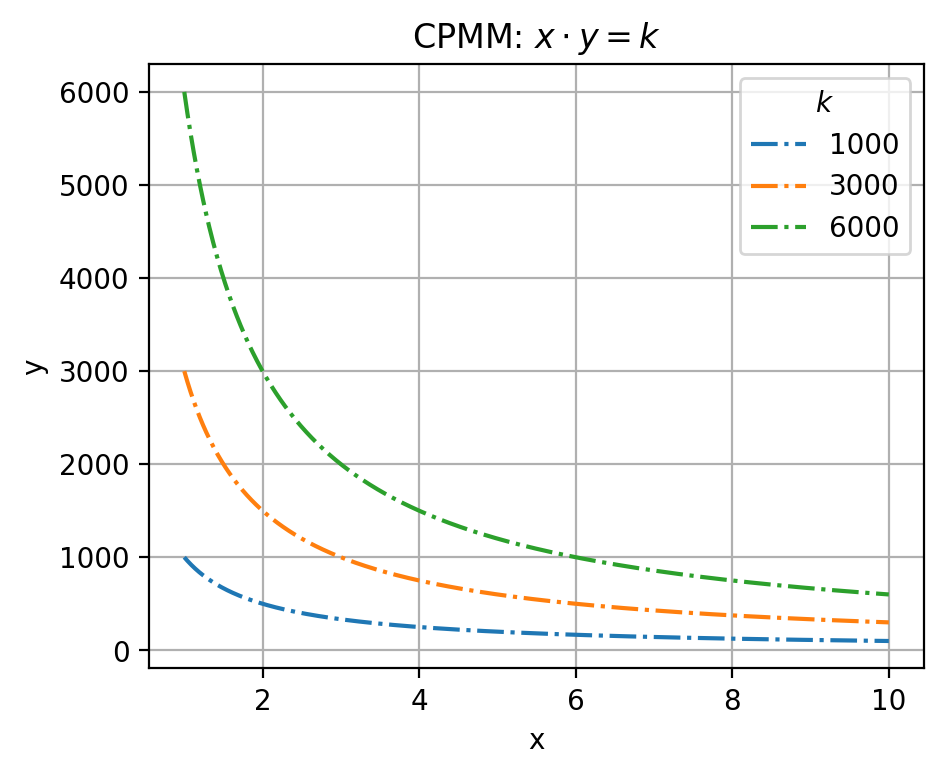}
  \caption{Trading function of a CPMM, such as Uniswap v2, for different levels of liquidity $k$.}
  \label{h2}
\end{subfigure}
\caption{\dhm{Samples of rectangular hyperbolae that define the set of possible states for different constant values of (a) temperature $T$ if the system is an ideal gas, (b) liquidity $k$ if we are considering a liquidity pool in Uniswap v2.}}
\label{fig:hyperbolae}
\end{figure}

\dhm{In thermodynamics, the system of interest is often an ideal gas. The state of this system can be specified by the pressure $P$ and volume $V$ that its particles respectively exert and fill, while its empirical temperature $T$ is measured via a test system kept in thermal equilibrium. Experimentally, a number of different systems are found to have an equation of state of the form
\begin{equation}
    PV = \phi(T),
\label{boyle}
\end{equation}
where $\phi$ is a general function with temperature $T$ as input variable \cite{thompson-stat-mech}. If the system is an ideal gas, then Eq. \eqref{boyle} unfolds into
\begin{equation}
    PV = nRT, 
\label{ideal-gas-law}
\end{equation}
which constraints the mutual evolution of variables $P, V$ and $T$, and is denoted as the \textit{ideal gas law}. The constant $n$ specifies the number of moles of particles in the considered closed system and $R\approx 8.31 \frac{J}{K \cdot mol}$ is the gas constant. 
For each fixed value of empirical temperature $T$, there is a well-defined set of possible states $(P,V)$ for the system in analysis. These states form a family of curves in the $(P,V)$ plane called \textit{isotherms}, which are rectangular hyperbolae as shown in Fig. \ref{h1}. Recalling the AMM Eq. \eqref{eq:AMM}, we notice how the trading function of a Constant Product Market Maker (CPMM) such as Uniswap v2 also defines a family of rectangular hyperbolae. These are proposed in Fig. \ref{h2}, where we sample curves by fixing different values of $k$ (i.e. liquidity) instead of temperature $T$. We have referred to Uniswap v2 for this comparison, since the concentrated liquidity mechanism of Uniswap v3 would require the introduction and discussion of a series of further details.}

\dhm{The similarity between fundamental relationships of the two systems mentioned, which is shown in Fig. \ref{fig:hyperbolae}, is the starting consideration that inspires our model.
Indeed, we propose an analogy in which each liquidity pool is a gas. Our intuition is that the amounts of reserves $(x,y)$, which define the state of a pool given some level of liquidity, can be enclosed into broader variables that govern all the dynamics relevant to a CPMM. Then, liquidity is approximated by computing the full TVL in each pool (i.e. our proxyTVL), since our data relate to Uniswap v3.
The view that we adopt is mainly macroscopic, and  especially based on resemblances between the thermodynamic explanation of the behaviour of gases and features of our crypto ecosystem. We have not developed a concrete full physical formulation of liquidity pools yet, but this is an in-progress extension of the current work.}

\dhm{Despite being very simple and elegant, the ideal gas law of Eq. \eqref{ideal-gas-law} properly captures  rich and interesting dynamics. To define our analogy and the similitude between variables (summarised in Table \ref{table:ideal-crypto-law}), we follow how the ideal gas law was discovered and reason about both the possible meaning of variables and expected relationships.}
First of all, $P$ can be intuitively compared to the USD volume traded by active LTs over e.g. a day.
If everything else but $T$ is kept constant, then we expect $T$ to increase with higher $P$. Therefore, we can indeed interpret $T$ as the liquidity of the pool, i.e. the value of our proxyTVL in USD for the pool at that date. 
Indeed, the evolution of liquidity of a pool accounts for the behaviour of LPs, and more LPs should execute mint operations when there are more LTs active, in order to collect higher profit from the fees that the latter pay for each swap transaction. Clearly, we also expect a stronger overall volume traded by LTs with more liquidity, due to more convenient, smaller price impact.
Variable $V$ is then the volume of the gas in the thermodynamics interpretation, i.e. ideal gas law. Despite being a more subtle relationship, we find that it is reasonable to consider $V$ as the stability of the exchange rate between the two tokens in the pool, i.e. $\textsc{std(Z)}^{-1}$, where $Z$ is the exchange rate.
Keeping everything else constant, one would expect that, with higher liquidity, the price is more stable. Similarly, a compressed gas with small volume will be less stable than an expanded gas. 
In addition to that, the \textit{concentrated liquidity} mechanism of Uniswap v3 implies that LPs encounter the risk of no gains from LTs' fees if the exchange rate moves outside the range of prices over which they are actively providing liquidity. Thus, a more stable $Z$ for the same USD volume traded by LTs (i.e. $P$) is likely to attract more minting operations, especially close to the current rate.
Finally, a higher $P$ with the same level of liquidity is indeed likely to cause a less stable relative price of the tokens, due to the impact of surplus swap operations.

Keeping the above in mind, Eq. \eqref{ideal-gas-law} thus becomes
\begin{equation}
\begin{split}
    P_{vol} \cdot V_{stab} &= n_{fee} \cdot R_{pool} \cdot T_{liq} \\
    \Rightarrow P_{vol} \cdot \textsc{std(Z)}^{-1} 
    &= feeTier^{-1} \cdot R_{pool} \cdot T_{liq},  \\
\end{split}
\label{ideal-crypto-law}
\end{equation}
allowing us to bring together into a single formula, all the variables that govern the three mechanisms of liquidity consumption, provision, and exchange rate evolution. 
Constant $R_{pool}$ is instead an invariant characteristic of each pool, \dhm{inferred from data.} Here, $n$ is the fixed number of moles (molecules), thus   a constant. A higher $n$ means more interactions \dhm{per unit of time} in the physical description, which we relate to lower feeTier and stronger activity of the LTs, that we know dominate LPs. Thus, we express $n=feeTier^{-1}$, which is indeed constant too.

\mc{  
\paragraph{Real gases and  van der Waals forces.} 
Our above model is based on the thermodynamics law for ideal gases. However, it is worth mentioning that  there also exists  a law for real gases that interact via van der Waals forces. This is governed by the  Van der Waals equation
\begin{equation}
    \Big( P+a\frac{n^2}{V^2} \Big) \cdot (V-nb) = nRT,
\end{equation}
where the variables have same meaning as before. In addition, $a$ is a constant whose value depends on the gas and represents intermolecular forces, while $b$ is the volume occupied by the molecules of one mole of the gas.
Based on a preliminary analysis, we believe it could be of interest to  expand the ideal crypto law in this direction. However, this is beyond the scope of our current work, and we leave this for future investigation.} 

\begin{table}[t]
\centering
\begin{tabular}[c]{ |p{1.5cm}|p{3cm}||p{3cm}|p{6cm}|  }
 \hline
 \rowcolor{LightCyan}
 \multicolumn{2}{|c||}{Ideal gas law} &\multicolumn{2}{|c|}{Ideal crypto law} \\
 \hline
 \cellcolor{Gray} Symbol & \cellcolor{LightGray} Meaning & \cellcolor{Gray} Symbol & \cellcolor{LightGray} Meaning\\
 \hline
 \hline
 $P$ & pressure & $P_{vol}$ & daily USD volume traded by LTs\\
 \hline
 $V$ & volume & $V_{stab}=\textsc{std(Z)}^{-1}$ & daily stability of the exchange rate $Z$\\
 \hline
 $n$ & moles of particles & $n_{fee} = feeTier^{-1}$ & stimulus to LTs' activation\\
 \hline
 $R$ & gas constant & $R_{pool}$ & pool crypto constant \\
 \hline
 $T$ & temperature & $T_{liq}$ & daily liquidity, i.e. proxyTVL value\\
 \hline
\end{tabular}
\vspace{0.2cm}
\caption{Parallelism drawn between the ideal gas law and our ideal crypto law.}
\label{table:ideal-crypto-law}
\end{table}


\subsection{Regression analysis and interpretation of results} 

As a first step, we test for empirical instances that support the validity of our ideal crypto law given by Eq. \eqref{ideal-crypto-law}.
We focus on case $A$ and consider the intersection of pools that are relevant for both liquidity provision and liquidity consumption. This results in a set of $16$ pools, namely USDC-USDT/100, WBTC-WETH/500, WETH-USDT/3000, WETH-ENS/3000, MATIC-WETH/3000, WBTC-USDC/3000, WBTC-WETH/3000, DAI-WETH/500, UNI-WETH/3000, SHIB-WETH/3000, USDC-USDT/500, USDC-WETH/500, WETH-CRV/10000, LINK-WETH/3000, USDC-WETH/3000, WETH-USDT/500.

\begin{figure}[t]
\centering
\begin{subfigure}[t]{.45\textwidth}
    \centering
    \includegraphics[width=0.9\linewidth]{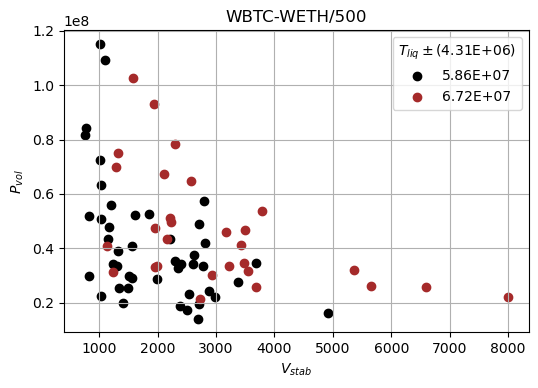}
      \label{fig:sub1-crypto-law}
      \caption{WBTC-WETH/500 pool.}
\end{subfigure}
\begin{subfigure}[t]{.45\textwidth}
    \centering
    \includegraphics[width=0.95\linewidth]{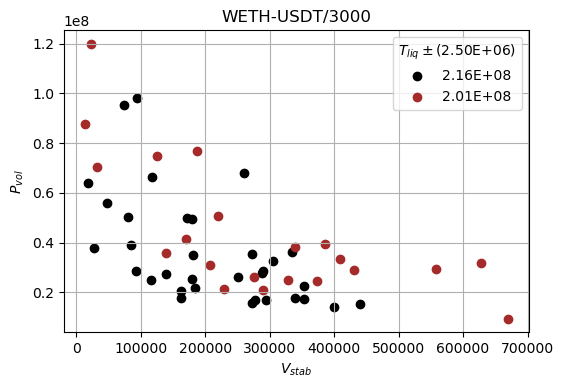}
      \label{fig:sub2-crypto-law}
      \caption{WETH-USDT/3000 pool.}
\end{subfigure}\vspace{0.3cm}\newline
\begin{subfigure}[t]{.45\textwidth}
    \centering
    \includegraphics[width=0.9\linewidth]{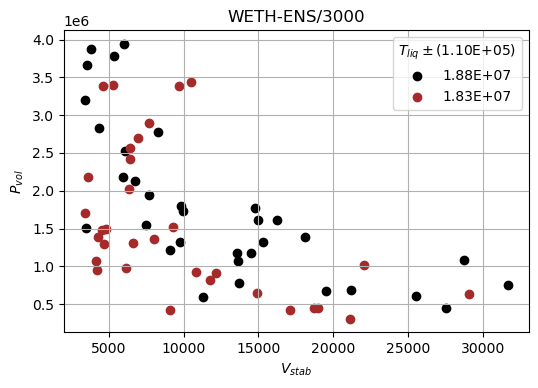}
      \label{fig:sub3-crypto-law}
      \caption{WETH-ENS/3000 pool.}
\end{subfigure}
\begin{subfigure}[t]{.45\textwidth}
    \centering
    \includegraphics[width=0.9\linewidth]{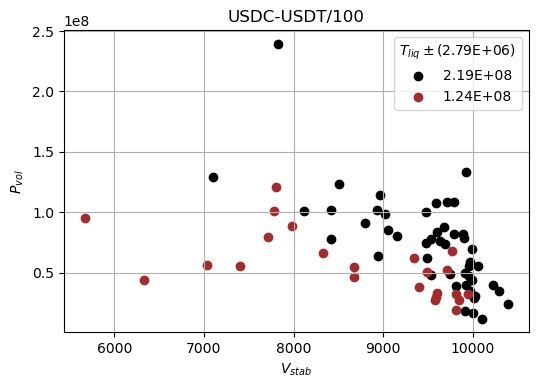}
      \label{fig:sub4-crypto-law}
      \caption{USDC-USDT/100 pool.}
\end{subfigure}   
\caption{\dhm{From the proposed ideal crypto law, we plot related empirical isotherms on the $(P_{vol}, V_{stab})$-plane for a sample of pools. We approximate data points at constant liquidity by measurements lying within one same small sub-range of proxyTVL, where these bins of proxyTVL are built to have equal width. We plot two isotherms per pool, in order to check for an upper shift in the curve related to a higher value of $T_{liq}$. Clearly, the scatter plots in (a), (b) and (c) do resemble the desired rectangular hyperbolae depicted in Fig. \ref{h1}, while (d) has no trace of the tested relationship. The latter result agrees with our intuition, since (d) concerns a pool of only stablecoins.}}
\label{fig:pairs-physics}
\end{figure}

For each individual pool, we compute the quantities $P_{vol}, V_{stab}, T_{liq}$ reported in Table \ref{table:ideal-crypto-law} at daily scale for the full six-months time window between January and June 2022. We suggest not to use higher frequencies due to significant general low rate of activity of LPs. 
\dhm{Observations with z-score $>3$ in absolute value are considered outliers and discarded. Then, the remaining daily realisations for each combination of pairs of variables are scatter-plotted to allow a critical analysis of the results. Figure \ref{fig:pairs-physics} shows a few representative examples, where we present data points related to isotherms on the $(P_{vol}, V_{stab})$-plane for four pools. We approximate data points at constant liquidity by measurements lying within one same small sub-range of proxyTVL, where these bins of proxyTVL are built to have equal width. Two isotherms per pool are plotted to check for an upper shift in the curve related to a higher value of $T_{liq}$.
We observe hints that the ansatz relationships are indeed satisfied, apart for the case of pools whose tokens are both stablecoins.} The latter dissimilarity is perhaps not surprising, since our ideal crypto law aims at encompassing the full ensemble of crypto mechanisms, while the behaviour of LTs \dhm{and price variation} on pools of stablecoins have lower relevance. 

\begin{figure}[t]
\centering
\begin{subfigure}{.45\textwidth}
    \centering
    \raisebox{4mm}[0pt][0pt]{
    \includegraphics[width=0.8\linewidth]{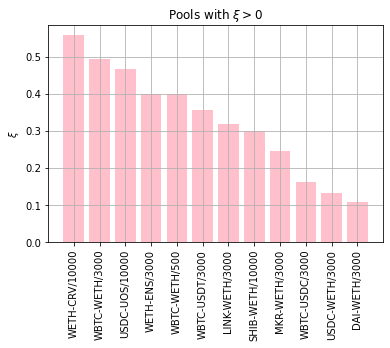}
    }
    \caption{Ranking of pools with cryptoness $\xi>0$.} 
      \label{fig:long-regression-a}
\end{subfigure}%
\begin{subfigure}{.50\textwidth}
    \centering
    \includegraphics[width=0.75\linewidth]{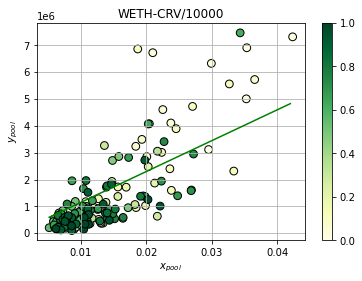}
    \caption{The $x_{pool},y_{pool}$ values for WETH-CRV/10000 pool, which has highest $\xi$. Dots are colour-coded to encode the temporal evolution, with lighter (resp. darker) ones denoting earlier (resp. more recent) observations. The straight line denotes the linear regression line.}   
    \label{fig:long-regression-b}
\end{subfigure}    
\caption{For each pool under analysis, the linear regression between $x_{pool},y_{pool}$ is computed for daily values over the six months of case $A$. Each related coefficient of determination provides the cryptoness $\xi$ of the pool.} 
\end{figure}

We broaden our sub-universe of venues of interest by taking the union of pools significant to LTs' and LPs' behaviour in case $A$. After \dm{filtering}
for relevance of the samples in the daily frequency, we are left with a set of $32$ pools.
For each pool, we perform a linear regression over the available six months of daily values, and thus estimate $R_{pool}$ by rearranging Eq. \eqref{ideal-crypto-law} to
\begin{equation}
\begin{split}
    P_{vol} &= R_{pool} \cdot \Big( \frac{ n_{fee} \cdot  T_{liq} }{V_{stab}} \Big) \\
    \Rightarrow y_{pool} 
    &= R_{pool} \cdot x_{pool},  \\
\end{split}
\label{rearrange}
\end{equation}
where the intercept is zero. We compute the coefficient of determination $R^2$ of the regression, which we refer to as the  \textit{cryptoness} $\xi$. Thus, pools with high cryptoness are meant to adhere well to our proposed ideal crypto law model.
We compute the average $\xi_{SS}$ over the seven pools found to exchange only stablecoins in our sub-universe, and compare it to the average $\xi_{notSS}$ of all other pools. We find values of $\xi_{SS} = -0.44$ and $\xi_{notSS} = -0.21$, meaning that SS pools should not be considered in our model, as expected and already motivated.
For the remaining pools, we show in Fig. \ref{fig:long-regression-a} the ones that have $\xi >0$.
We notice that different feeTiers appear to be relevant and that there is one interesting occurrence of two pools, both with high $\xi$, that exchange the same tokens, i.e. WBTC-WETH/3000 and WBTC-WETH/500. This latter result raises the question of whether these two pools are generally adhering to our ideal crypto law, or if they might follow it at disjoint periods of time that however influence the overall cryptoness values and render them both significant. Before investigating this idea further, it is worth  highlighting the fact that several pools with high $\xi$ are seen to exchange exotic tokens.
Examples are pools trading CRV and UOS, which are respectively the tokens of the well-know DEX Curve.fi, and of the blockchain-based gaming company Ultra.
The general pattern that we read is that pools exchanging digital assets linked to a tangible and established business idea might adhere better to our ideal crypto law.
We think of such pools\dhm{, and companies, as being strongly embedded} in the crypto ecosystem, at least over the time window of reference in our  analyses.

For the sake of clarity, we explicitly depict in Fig. \ref{fig:long-regression-b} the linear regression pursued for the pool with highest cryptoness, i.e. WETH-CRV/10000, where occurrences that happened more recently in time are color-coded darker.
We also mention that pools might exhibit  a less good fit to our ideal crypto law due to noise introduced by the characteristic frequency of executed swaps, mint and burn operations on each pool.
Liquidity provision events are generally rare, as reported in Table \ref{table:frequencies}. Thus, it might become necessary to define ad-hoc frequencies, possibly dynamic, to be used in the regression for each pool, in order to better investigate the related evolution of behaviours.

\begin{table}[h]
\centering
\begin{tabular}[c]{ |p{3cm}|p{1.5cm}||p{3cm}|p{1.5cm}|p{3cm}|p{1.5cm}|  }
 \hline
 \rowcolor{LightCyan}
 \multicolumn{2}{|c|}{Liquidity consumption} &\multicolumn{4}{|c|}{Liquidity provision} \\
 \hline
 \cellcolor{Gray} Pool & \cellcolor{LightGray} Daily Swap & \cellcolor{Gray} Pool & \cellcolor{LightGray} Daily Mint & \cellcolor{Gray} Pool &\cellcolor{LightGray} Daily Burn\\
 \hline
 \hline
 \rowcolor{green}
 USDC-WETH/500 & 6181 & USDC-WETH/3000 & 95 & USDC-WETH/3000 & 70\\
 \hline
 \rowcolor{green}
 WETH-USDT/500 & 2395 & USDC-WETH/500 & 65 & USDC-WETH/500 & 70\\
 \hline
 \rowcolor{green}
 DAI-WETH/500 & 817 & WBTC-WETH/3000 & 19 & WBTC-WETH/3000 & 26\\
 \hline
 \hline
 \rowcolor{orange}
 WETH-CRV/10000 & 55 & DYDX-WETH/3000 & 1.0 & FXS-WETH/10000 & 1.4\\
 \hline
 \rowcolor{orange}
 USDC-UOS/10000 & 39 & USDC-UOS/10000 & 0.6 & USDC-UOS/10000 & 1.0\\
 \hline
 \rowcolor{orange}
 SHIB-WETH/10000 & 32 & CEL-WETH/3000 & 0.4 & CEL-WETH/3000 & 0.8\\
 \hline
\end{tabular}
\vspace{0.2cm}
\caption{Average daily frequency of operations on different pools for the time window of case $A$. We report the first and last three values when sorting by magnitude.}
\label{table:frequencies}
\end{table}

\begin{figure}[t]
\centering
\begin{subfigure}{.5\textwidth}
    \centering
    \includegraphics[width=0.9\linewidth]{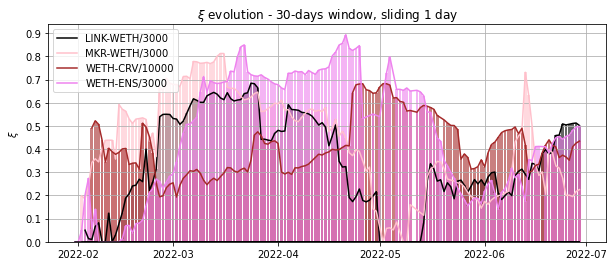}
    \caption{Evolution of $\xi$ for a sample of pools exchanging exotic tokens.}
      \label{fig:exotic-reg}
\end{subfigure}%
\begin{subfigure}{.5\textwidth}
    \centering
    \includegraphics[width=0.9\linewidth]{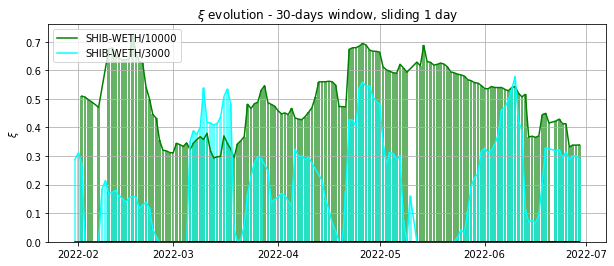}
    \caption{Evolution of $\xi$ for two pools exchanging SHIB-WETH.}
      \label{fig:shib-reg}
\end{subfigure}    
\vspace{0.3cm}
\begin{subfigure}{.6\textwidth}
    \centering
    \includegraphics[width=\linewidth]{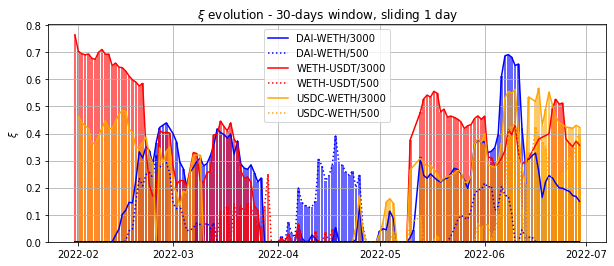}
    \captionsetup{width=1.2\textwidth}
    \caption{Evolution of $\xi$ for a sample of pools exchanging WETH vs. stablecoin. The dominance of different pools over time can lead to hypothesise cryptoness as a measure of the varying health of pools and indicator of where market participants should prefer to be active on.}
      \label{fig:weth-reg}
\end{subfigure}  
\caption{Evolution of cryptoness $\xi$ for different subsets of pools, where $\xi$ is now repeatedly computed from the regression over a 30-days window of instances, sliding one day each time.}
\label{fig:physics-temporal-regressions}
\end{figure}

\paragraph{\dhm{Dynamic} analyses.} We now analyse the evolution of our proposed  cryptoness metric for the set of pools, over the six-months time window of case $A$, i.e. Jan-June 2022. \mc{We again compute the regression in Eq. \eqref{rearrange} and record the related coefficient of determination as the cryptoness value $\xi$, but we now use the observations  contained in a $30$-day window in time, sliding every day.
This allows us to recognise pools that adhere more or less to our ideal crypto law at different points in time, and investigate the patterns generated. Interesting results for specific subsets of pools are shown in Fig. \ref{fig:physics-temporal-regressions}, where we threshold all the irrelevant $\xi < 0$ 
to $\xi = 0$ for ease of visualisation.
In reference to our ranking of Fig. \ref{fig:long-regression-a} and related discussion, we consider in Fig. \ref{fig:exotic-reg} the evolution of $\xi$ for a subset of pools exchanging exotic tokens, which 
relate to well-established blockchain-based companies. As expected, we observe $\xi$s that are strongly significant for almost the entire six-months time range under consideration. Thus, we confirm that the exotic tokens depicted are deemed by our model to have solid associated companies by the dynamics of market participants and price, and we construe the related pools as \textit{healthy} venues for trading. }

\dm{Our intuition then suggests that there should exist \textit{only one ideal feeTier per point in time, for the exchange of two same tokens}. This feeTier would be the key that balances the set of mechanisms we are modeling and the beliefs of the market participants. 
As a first related example, Fig. \ref{fig:shib-reg} shows some association between the drops in cryptoness of the liquidity pool SHIB-WETH/10000, and the peaks in cryptoness of SHIB-WETH/3000, allowing us to consider our cryptoness measure as an indicator of the described tendency.
With a similar view, we also expect
the redundancy of pools exchanging the same token against different stablecoins to reduce the overall health of single pools, due to the related fragmentation of market participants' dynamics. 
Thus, we believe that a natural stabilisation of each token to one specific stablecoin per period in time should also be reached with the stronger establishment of crypto ecosystems.}

\begin{figure}[h]
    \begin{subfigure}{.4\textwidth}
      \centering
      \includegraphics[width=.99\linewidth]{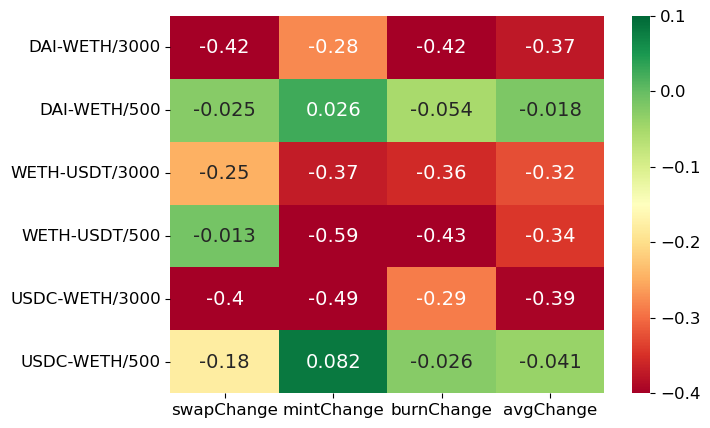}
      \caption{Variation in activity, i.e. in the frequency of the different transaction types, over April.} 
      \label{fig:physics-heatmap}
    \end{subfigure}%
    \begin{subfigure}{.6\textwidth}
      \centering
      \includegraphics[width=.99\linewidth]{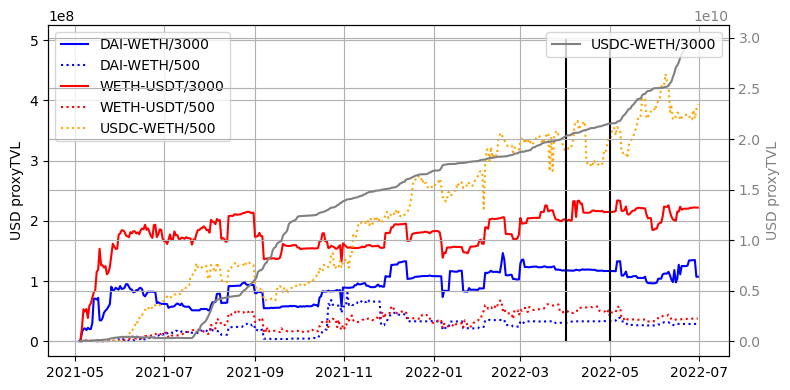}
      \caption{Our proxyTVL in USD.}
      \label{fig:physics-tvl}
    \end{subfigure}
    \caption{Comparison of activity variation and liquidity evolution between our six pools exchanging WETH against a stablecoin, over the six months of case $A$. We discuss changes in average activity over April 2022 versus on the remaining five months of case $A$, and similarly compare disequilibria of minting versus burning operations in USD traded.     Pool DAI-WETH/500 is the only venue of the subset with significant cryptoness in April 2022 and indeed, it reports both the least damage in activity and no drop in proxyTVL over that month.}
    \label{fig:physics-final}
\end{figure}

\dm{The results of this section provide supporting evidence towards the above concepts. In particular, we show that our cryptoness measure signals the healthiest venues (i.e. liquidity pools) where agents should be active on, by comparing simultaneous changes in $\xi$, versus variations in agents' activity and TVL. This is performed for the set of relevant pools that exchange WETH against stablecoins, i.e. DAI-WETH, WETH-USDT and USDC-WETH, also with different feeTiers, i.e. 500 and 3000.} 
Figure \ref{fig:weth-reg} shows the related evolutions of the cryptoness measure, where we can clearly see that feeTier 3000 is the relevant one for each pair DAI-WETH, WETH-USDT and USDC-WETH for the \mc{entire time window, except during April 2022.  Interestingly, over this month, no liquidity pools exhibit  strong cryptoness, except for  DAI-WETH/500, which we thus claim to be the only healthy venue for trading at that point in time. 
To support our claim}, we compute the average daily number of swap actions, mint operations and burn operations, for each one of our six pools under analysis, as related measures of activity and good usage. 
In particular, we compute average quantities over the month of April, and then over the time window Jan-June 2022 but excluding April, and then calculate the percentage change. This results in the set of values
\begin{equation}
    opChange = \frac{opApril-opNotApril}{opNotApril},
\end{equation}
where \say{op} indicates an operation between swap, mint and burn. The actual computations are reported in Fig. \ref{fig:physics-heatmap}, where we further include \textit{avgChange} as the average of \textit{swapChange}, \textit{mintChange} and \textit{burnChange}. 
In addition, we also plot in Fig. \ref{fig:physics-tvl} the evolution of TVL in USD over the pools of interest, in order to be able to compare occurrences of drops in liquidity over time.
Figure \ref{fig:physics-heatmap} reveals that DAI-WETH/500 is indeed the liquidity pool with the least damage of activity during April 2022. While there exists another pool which performs better than others, namely USDC-WETH/500, it suffered from a significant drop in liquidity in April 2022. On the other hand, DAI-WETH/500 had constant TVL during this month, as clearly shown in Fig.  \ref{fig:physics-tvl}.
\mc{We conclude that the only pool with a significant cryptoness $\xi$ score during April 2022 is indeed the healthiest and preferred trading venue  during the month of relevance. This provides further empirical motivation and utility} to our proposed ideal crypto law. 

\begin{figure}[h]
\centering
\begin{subfigure}{.33\textwidth}
    \centering
    \includegraphics[width=0.95\linewidth]{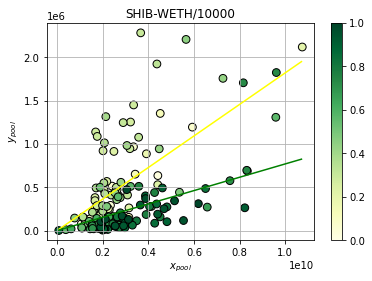}
    \caption{SHIB-WETH/10000 pool.}
      \label{fig:r2a}
\end{subfigure}%
\begin{subfigure}{.32\textwidth}
    \centering
    \includegraphics[width=0.95\linewidth]{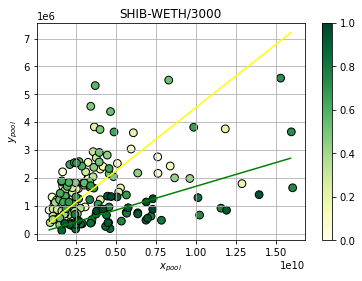}
    \caption{SHIB-WETH/3000 pool.}
      \label{fig:r2b}
\end{subfigure}%
\begin{subfigure}{.33\textwidth}
    \centering
    \includegraphics[width=0.95\linewidth]{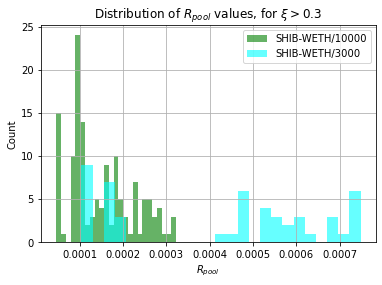}
    \caption{Distribution of $R_{pool}$'s.}
      \label{fig:dist-r-pool}
\end{subfigure}  
\caption{\dhm{Evolution in time of the characteristic coefficient $R_{pool}$ for two pools exchanging the same tokens SHIB and WETH, but with feeTier $10000$ and $3000$. In (a) and (b), the yellow (green) line is generated from the regression over data from the first four months (last two months) of case study $A$. In (c), we propose the distribution of $R_{pool}$ values recovered from the dynamic analyses with reference to Fig. \ref{fig:shib-reg}. To lower the noise level, we keep only values of $R_{pool}$ when the related cryptoness is $\xi>0.3$.}}
\label{fig:evol-r-pool}
\end{figure}

Our measure of cryptoness of liquidity pools aims at becoming a useful tool for market participants, allowing them to assess the health of trading venues and decide the best environments on which to be active. 
Ideally, we would like to witness smooth dynamics of the evolution of cryptoness in the future, with new pools adhering better to the ideal crypto law once they 
become well-rooted components of the crypto ecosystem.
From a \mc{regulatory} point of view, 
dynamic requirements could then be imposed, such as less over-collateral required for trades on pools with cryptoness above a certain threshold. 
\mc{This metric could thus be employed both by regulators and practitioners for developing pool health monitoring tools, and establishing minimum
levels of requirement.}  
\dhm{Finally, we believe that it would also be interesting to investigate and interpret variations in the characteristic constant $R_{pool}$ for specific pools, especially after drops in cryptoness. As a basic motivating example, Figs. \ref{fig:r2a} and \ref{fig:r2b} show two regressions over the first four and final two months of case $A$, for pools SHIB/WETH-10000 and SHIB/WETH-3000. From the plots, it is indeed clear that the slope of regression lines varies significantly.
The related distributions of recovered $R_{pool}$ values for a sliding window long $30$-days are then shown in Fig. \ref{fig:dist-r-pool}, where we are leveraging on the results from the dynamic analyses on SHIB/WETH pools that generated Fig. \ref{fig:shib-reg} too. We keep only $R_{pool}$ values related to regressions with $\xi > 0.3$, in order to lower noise. Interestingly, the two pools have $R_{pool}$ values with same order of magnitude, which is a general feature recovered from pools exchanging same tokens.}


\section{Conclusions}
\label{sec:conclusions}

Blockchain, DeFi and DEXs are recent concepts that just started taking roots in the common language and knowledge of \mc{both practitioners and academics}. However, a real comprehension of  the characteristic dynamics of these protocols is still far away. Similarly, academic research is at its dawn on the topic, despite having strongly accelerated in the past year.
\mc{To this end, our investigations aim at being a  stepping stone towards a deeper understanding of the crypto ecosystem, and we achieve this task} by empirically studying and characterising Uniswap v3 DEX. 
We build a workflow to define the most relevant liquidity pools over time by assessing the inner features of pools along with their interconnectedness, and provide related \dm{lists of liquidity pools significant for six different windows in time, i.e. cases $A/B1/B2/C1/C2/C3$,} 
that can be directly used for future research studies. We then focus on LTs and show the existence of seven \say{species} of traders with interpretable features. These clusters are recovered by assessing the equivalence of LTs structural trading behaviour, and suggest a connection between patterns in swap transactions and specific types of pools on which these operations are indeed executed.
Finally, we also propose a novel metric that could aid practitioners and regulators in the challenge of assessing the \say{health} of different trading venues, i.e. liquidity pools, by proposing an ideal crypto law and proving the efficacy of the related cryptoness goodness measure.

\paragraph{Future work.}  
Regarding future directions of research, there are three 
main  threads we aim to pursue.
The first one is a detailed investigation into the behaviour of LPs, \dm{along with the corresponding clustering of species, also} leveraging on a strongly data-driven approach.   
\dm{Secondly}, we are considering the development of a broader \textit{integrated crypto market indicator}\dm{. This would be} based on an  aggregated  
cryptoness measure over a set of pools, \mc{with different associated weights quantifying the contribution to the index, proportional to} the evolving total value locked in each liquidity pool. And \dm{as a last point}, we plan to leverage
the entire data on the Ethereum blockchain to track the flow of funds over multiple DEXs and active protocols. This will \mc{enable us to gain a} better understanding of LTs, as we will then be able to approximate their profit and loss (PnL). Indeed, the crypto ecosystem is highly interconnected, and agents easily trade between different exchanges on the same blockchain, also with the possibility to enhance their positions via borrowing. 
In parallel, one could also investigate the \say{optimal routing problem} \cite{routing} on the Ethereum blockchain, which is formulated as the problem of optimally executing an order involving multiple crypto assets on a network of multiple constant function market makers.

\section*{Acknowledgements} 
\vspace{-3mm}  
We are grateful to \'{A}lvaro Cartea, Fay\c{c}al Drissi, Marcello Monga and Andrea Pizzoferrato for insightful discussions. Deborah Miori acknowledges financial support from the \textit{EPSRC CDT in Mathematics of Random Systems} (EPSRC Grant  EP/S023925/1).

\section*{Statements and Declarations} 
\vspace{-3mm}  
The authors have no competing interests to declare that are relevant to the content of this article.


\bibliographystyle{unsrt}  
\bibliography{references}

\appendix

\section{Sub-universes of pools for cases B1/B2/C1/C2/C3}
\label{appendix:A}

We report here the pools that our methodology finds most significant for cases $B1/B2/C1/C2/C3$. For each one of these time windows, we list separately the pools relevant for liquidity consumption (LT data) and liquidity provision (LP data) investigations. We also provide the thresholds chosen to define the giant components in our interconnectedness analyses, always with reference to the workflow of Section \ref{sec:pools-of-interest}.

\paragraph{Case $B1$.} LT data: thresholds for common origins, senders and bridges are $1,500$ and $60$ and $600$ respectively. We recover a final set of $32$ pools, which are: DAI-WETH/500, FRAX-USDC/500, SPELL-WETH/3000, agEUR-USDC/500, USDC-USDT/100, LUSD-USDC/500, USDC-UST/100, WBTC-WETH/3000, DAI-USDC/100, USDC-NCR/500, WETH-USDT/500, XSGD-USDC/500, LINK-WETH/3000, USDC-WETH/3000, DAI-WETH/3000, WETH-BTRFLY/10000, FEI-USDC/500, HEX-USDC/3000, WETH-ENS/3000, MATIC-WETH/3000, USDC-USDT/500, XSGD-WETH/500, WBTC-WETH/500, FXS-WETH/10000, GALA-WETH/3000, WBTC-USDC/3000, WETH-CRV/10000, WETH-USDT/3000, USDC-UOS/10000, HEX-WETH/3000, USDC-WETH/500, USDC-GF/3000.

LP data: thresholds for common origins and senders are $20$ and $3$ respectively. We recover a final set of $16$ pools, which are: DAI-WETH/500, WBTC-USDC/3000, USDC-WETH/3000, LINK-WETH/3000, WETH-CRV/10000, WBTC-WETH/3000, MATIC-WETH/3000, WETH-ENS/3000, USDC-USDT/100, UNI-WETH/3000, SHIB-WETH/3000, USDC-WETH/500, WETH-USDT/500, MKR-WETH/3000, SHIB-WETH/10000, WBTC-WETH/500.

\paragraph{Case $B2$.} LT data: thresholds for common origins, senders and bridges are $1,500$ and $80$ and $600$ respectively. We recover a final set of $33$ pools, which are: DAI-WETH/500, FRAX-USDC/500, FEI-USDC/100, USDC-USDT/100, UST-WETH/3000, HDRN-USDC/10000, USDC-STG/3000, CEL-WETH/3000, DAI-USDC/500, WBTC-WETH/3000, DAI-USDC/100, WETH-USDT/500, APE-WETH/3000, LINK-WETH/3000, USDC-WETH/3000, DAI-WETH/3000, USDC-WETH/10000, HEX-USDC/3000, WETH-ENS/3000, MATIC-WETH/3000, USDC-USDT/500, APE-USDC/3000, WBTC-WETH/500, WBTC-USDC/3000, WETH-USDT/3000, WETH-LUNA/10000, USDC-UOS/10000, BUSD-USDC/500, HEX-WETH/3000, WETH-LOOKS/3000, SHIB-WETH/3000, USDC-WETH/500, DAI-FRAX/500.

LP data: thresholds for common origins and senders are $15$ and $3$ respectively. We recover a final set of $16$ pools, which are: WBTC-USDT/3000, DAI-WETH/500, WBTC-USDC/3000, LINK-WETH/3000, USDC-WETH/3000, WETH-USDT/3000, WETH-LUNA/10000, HEX-USDC/3000, WBTC-WETH/3000, MATIC-WETH/3000, USDC-USDT/500, HEX-WETH/3000, UNI-WETH/3000, USDC-WETH/500, WETH-USDT/500, SHIB-WETH/10000.

\paragraph{Case $C1$.} LT data: thresholds for common origins, senders and bridges are $1,000$ and $50$ and $400$ respectively. We recover a final set of $29$ pools, which are: DAI-WETH/500, FRAX-USDC/500, SPELL-WETH/3000, agEUR-USDC/500, USDC-USDT/100, USDC-UST/100, WBTC-WETH/3000, DAI-USDC/100, USDC-NCR/500, WETH-USDT/500, XSGD-USDC/500, LINK-WETH/3000, USDC-WETH/3000, DAI-WETH/3000, WETH-BTRFLY/10000, FEI-USDC/500, HEX-USDC/3000, WETH-ENS/3000, MATIC-WETH/3000, SOS-WETH/10000, XSGD-WETH/500, WBTC-WETH/500, FXS-WETH/10000, GALA-WETH/3000, WBTC-USDC/3000, WETH-CRV/10000, WETH-USDT/3000, HEX-WETH/3000, USDC-WETH/500.

LP data: thresholds for common origins and senders are $10$ and $3$ respectively. We recover a final set of $15$ pools, which are: DAI-WETH/500, WBTC-USDC/3000, LINK-WETH/3000, USDC-WETH/3000, WETH-CRV/10000, WETH-BTRFLY/10000, WBTC-WETH/3000, WETH-ENS/3000, MATIC-WETH/3000, UNI-WETH/3000, USDC-WETH/500, SHIB-WETH/3000, WETH-USDT/500, MKR-WETH/3000, SHIB-WETH/10000.

\paragraph{Case $C2$.} 
LT data: thresholds for common origins, senders and bridges are $1,000$ and $50$ and $500$ respectively. We recover a final set of $30$ pools, which are: DAI-WETH/500, FRAX-USDC/500, WETH-WRLD/10000, USDC-USDT/100, DAI-USDC/500, USDC-UST/100, WBTC-WETH/3000, DAI-USDC/100, USDC-NCR/500, WETH-USDT/500, XSGD-USDC/500, LINK-WETH/3000, USDC-WETH/3000, DAI-WETH/3000, WETH-BTRFLY/10000, HEX-USDC/3000, WETH-ENS/3000, MATIC-WETH/3000, XSGD-WETH/500, WBTC-WETH/500, FXS-WETH/10000, GALA-WETH/3000, WBTC-USDC/3000, WETH-USDT/3000, HEX-WETH/3000, USDC-RSS3/3000, WETH-LOOKS/3000, SHIB-WETH/3000, USDC-WETH/500, DAI-FRAX/500.

LP data: thresholds for common origins and senders are $10$ and $3$ respectively. We recover a final set of $16$ pools, which are: DAI-WETH/500, WBTC-USDC/3000, LINK-WETH/3000, USDC-WETH/3000, WETH-USDT/3000, WETH-LUNA/10000, WETH-WRLD/10000, WBTC-WETH/3000, MATIC-WETH/3000, USDC-USDT/100, UNI-WETH/3000, SHIB-WETH/3000, USDC-WETH/500, WETH-USDT/500, MKR-WETH/3000, SHIB-WETH/10000.

\paragraph{Case $C3$.} 
LT data: thresholds for common origins, senders and bridges are $1,000$ and $75$ and $500$ respectively. We recover a final set of $30$ pools, which are: DAI-WETH/500, FRAX-USDC/500, FEI-USDC/100, USDC-USDT/100, UST-WETH/3000, HDRN-USDC/10000, CEL-WETH/3000, WBTC-WETH/3000, DAI-USDC/100, WETH-USDT/500, UNI-WETH/3000, APE-WETH/3000, LINK-WETH/3000, USDC-WETH/3000, DAI-WETH/3000, HEX-USDC/3000, WETH-ENS/3000, MATIC-WETH/3000, USDC-USDT/500, APE-USDC/3000, WBTC-WETH/500, WBTC-USDC/3000, WETH-CRV/10000, WETH-USDT/3000, WETH-LUNA/10000, BUSD-USDC/500, HEX-WETH/3000, WETH-LOOKS/3000, SHIB-WETH/3000, USDC-WETH/500.

LP data: thresholds for common origins and senders are $10$ and $3$ respectively. We recover a final set of $11$ pools, which are: WBTC-USDT/3000, DAI-WETH/500, WBTC-USDC/3000, USDC-WETH/3000, UNI-USDC/3000, WETH-USDT/3000, WETH-LUNA/10000, HEX-USDC/3000, HEX-WETH/3000, WETH-USDT/500, USDC-WETH/500.

\end{document}